\documentclass[aps,pre,reprint,showpacs]{revtex4-1}

\usepackage[utf8x]{inputenc}
\usepackage[T1]{fontenc}
\usepackage[british, english]{babel}

\usepackage{amstext}
\usepackage{amsmath}
\usepackage{amssymb}
\usepackage{amsfonts}

\usepackage{graphicx}

\usepackage[
	pdftitle={Buckling of Spherical Capsules},
         pdfauthor={Sebastian Knoche, Jan Kierfeld}
         ]{hyperref}

\newcommand*{\diff}{\mathrm{d}}			
        	
\newcommand{\del}{\partial}

\begin{document}

\title{Buckling of Spherical Capsules} 
\author{Sebastian Knoche}
\email{sebastian.knoche@tu-dortmund.de} 
\author{Jan Kierfeld}
\email{jan.kierfeld@tu-dortmund.de} 
\affiliation{Department of Physics,
  Technische Universit\"{a}t Dortmund, 44221 Dortmund, Germany}


\begin{abstract}
  We investigate buckling of soft elastic capsules under negative pressure 
  or  for reduced capsule volume. 
  Based on nonlinear shell theory and the assumption
  of a hyperelastic capsule membrane, shape equations for axisymmetric and
  initially spherical capsules are derived and solved numerically. A
  rich bifurcation behavior is found, which is presented in terms of
  bifurcation diagrams. The energetically preferred stable configuration is
  deduced from a least-energy principle both for prescribed volume and
  prescribed pressure.  
  We find that buckled shapes are energetically favorable already at smaller 
  negative pressures and larger critical volumes 
  than predicted by the classical buckling instability. 
  By preventing self-intersection for strongly reduced
  volume, we obtain a complete picture of the buckling process and can follow
  the shape from the initial undeformed state through the buckling instability
  into the fully collapsed state. 
  Interestingly, the sequences of bifurcations
  and stable capsule shapes differ for prescribed volume and prescribed
  pressure. In the buckled state, we find a relation between curvatures at the
  indentation rim and the bending modulus, which can be used to determine
   elastic moduli  from experimental shape analysis.
\end{abstract}

\maketitle

\section{Introduction}

An elastic capsule consists of an elastic membrane enclosing a fluid
phase. Elastic capsules are commonly met in nature, prominent examples are red
blood cells or virus capsules. Artificial capsules can be fabricated by
various methods \cite{Meier2000}, for example by interfacial polymerization at
liquid droplets \cite{Rehage2002} or by multilayer deposition of
polyelectrolytes \cite{Donath1998}, and have numerous applications. Sizes of
capsules vary from the nanometer to the micrometer regime, and their
mechanical properties depend on the fabrication process. For various
applications, for example, if capsules are used as delivery and release
systems, there is a need to characterize mechanical properties of
capsules. The shape of capsules is approximately spherical but capsules are
easily deformed by shear flow \cite{Barthes-Biesel2011}, rotation
\cite{Pieper1998}, in adhesion \cite{Komura2005,Graf2006}, 
or by the application of local
forces \cite{Fery2004,Fery2007,Zoldesi2008}. Their deformation behavior also
exhibits buckling instabilities upon decreasing the interior pressure or the
enclosed volume \cite{Gao2001,Fery2004,Zoldesi2008,Sacanna2011}
or in adhesion \cite{Komura2005}. All these deformation
modes can potentially be used in experiments to infer material properties of
capsules. 

Changes of the pressure inside the capsule by osmosis or mechanical means and
changes in the capsule volume represent the most basic deformation mechanisms
for capsules with spherical rest shape. In this article, we study the collapse
of a three-dimensional spherical capsule via the buckling instability into a
fully collapsed state under negative pressure or  for reduced capsule 
volume. The mechanical 
buckling instability sets in at the classical buckling pressure, which is
well-known within linear shell theory for small displacements and an isotropic
material \cite{LL7} and has recently been extended to shell materials with
anisotropic shear response \cite{Ru2009}. The classical buckling pressure
$p_{cb}\propto -E(H_0 /R_0)^2$, where $E$ is Young's modulus, $H_0$ the initial
membrane thickness and $R_0$ its initial diameter, can be used in experiments
to determine Young's modulus of the capsule material \cite{Gao2001,Fery2004}.

While the classical buckling pressure only marks the onset of the instability
within linear shell theory and, thus, the limits of stability of a spherical
capsule, it is much more difficult to calculate buckled shapes beyond the
critical buckling pressure because larger deformations require nonlinear
theories and contact between originally opposite capsule sides has to be
included in order to prevent self-intersections. Some nonlinear theories have
been applied to compute axisymmetric shapes with large deformations, for
example in \cite{Blyth2004} under the assumption of isotropic tensions and for
a more general case in \cite{Bauer1970}. 
Large buckling deformations have been considered numerically 
based on triangulated surface models 
\cite{Komura2005,Vliegenthart2011,Quilliet2008,Marmottant2011}.
In the framework of shell theory, however, a complete picture of the
transition into buckled shapes as observed in the experiments is still
lacking. 

In order to develop this picture for axisymmetric capsules, we use a nonlinear
shell theory \cite{Libai1998,Pozrikidis2003} and assume hyperelastic capsule
membranes. For hyperelastic materials, a strain-energy function exists from
which the tensions and bending moments can be deduced. It can be shown in
general that solutions of the equations of force and moment equilibrium render
the functional of total energy stationary \cite{Libai1998}. This allows us to
combine two tools, force equilibrium and principle of minimal energy, in order
to find different branches of stationary deformed capsule shapes and then
determine which capsule shape among different branches represents the global
energy minimum. At the classical buckling instability the branch corresponding
to a spherical capsule loses stability and a bifurcation to buckled shapes
takes place. This bifurcation is analyzed in detail beyond a linear stability
analysis and in energy (and enthalpy) bifurcation diagrams.  We find that the
buckling transition is discontinuous in the energy diagrams and that the
buckled shape becomes {\em energetically} favorable already at a smaller
negative pressure $|p_c|<|p_{cb}|$ than the classical buckling pressure, where
the spherical shape becomes mechanically unstable.

We extend our approach to capsules with opposite sides in 
contact in order to prevent self-intersection at strongly reduced
capsule volume. As far as we know, this has so far only been achieved for
elastic rings \cite{Flaherty1972} in two dimensions, but not for spherical
shells in three dimensions.

Furthermore, we analyze features of buckled shapes, in particular, the maximal
curvature occurring at the edge of 
 the indentation rim and find a relation between
the maximal curvatures and the bending modulus, which can be used in
experimental shape analysis.

\section{Finite Strain Shell Theory}

\subsection{Geometric Setup}

\begin{figure}[tb]
  \centerline{\includegraphics[width=86mm]{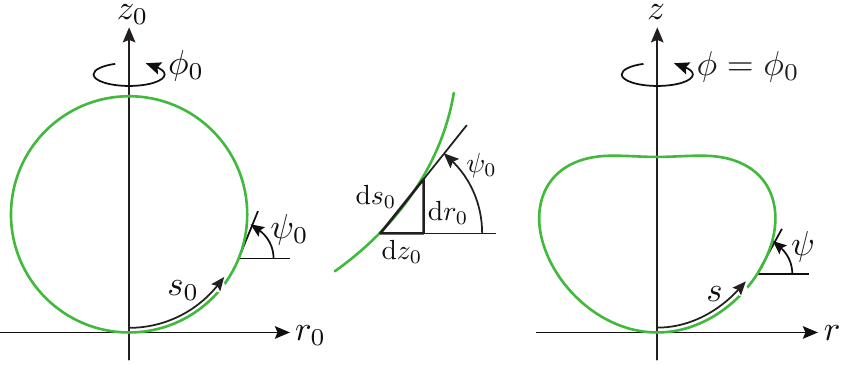}}
  \caption{(Color online) Parameterization of the undeformed (left; always with
    index ``0'') and deformed midsurface}
\label{fig:midsurface}
\end{figure}

We consider axisymmetric capsules, which are 
 undergoing axisymmetric, torsionless
deformations and whose membrane thickness is small compared to the other
capsule dimensions. The midsurface of the undeformed capsule is parameterized
by the curvilinear coordinates $s_0 \in [0,\, L_0]$ and $\phi_0 \in [0,\,
2\pi)$ denoting the arc length measured along the meridian and the angle of
revolution, respectively. Its shape is determined by the functions $r_0(s_0)$
and $z_0(s_0)$ in cylindrical polar coordinates (see figure
\ref{fig:midsurface}).

In addition to the functions $r_0(s_0)$ and $z_0(s_0)$, the slope angle
$\psi_0(s_0)$ is defined via the relations (see figure \ref{fig:midsurface})
\begin{equation}
  \frac{\diff r_0}{\diff s_0} = \cos \psi_0 \quad \text{and} \quad 
  \frac{\diff z_0}{\diff s_0} = \sin \psi_0. 
\label{eq:geom1}
\end{equation}
Using this parameterization we calculate the \emph{principal curvatures} of
the midsurface,
\begin{equation}
 \kappa_{s_0} = \frac{\diff \psi_0}{\diff s_0} \quad \text{and} \quad 
 \kappa_{\phi_0} = \frac{\sin \psi_0}{r_0} 
\label{eq:geom2}.
\end{equation} 
Here, $\kappa_{s_0}$ denotes the curvature in the meridional direction, and
$\kappa_{\phi_0}$ the curvature in the circumferential direction.

For the special case of capsules with spherical resting shape, the
parameterization of the undeformed midsurface is known analytically,
\begin{equation}
\begin{aligned}
 r_0(s_0) &= R_0 \, \sin\left( s_0 / R_0 \right) \quad \text{and} \\
 z_0(s_0) &= R_0 \, \left[1- \cos\left( s_0 / R_0 \right)\right],
\label{eq:undef1}
\end{aligned}
\end{equation} 
where the arc length $s_0$ ranges up to $L_0 = \pi \, R_0$. 
Accordingly, the slope angle and curvatures are given by
\begin{equation}
  \psi_0(s_0) = {s_0 / R_0} \quad \text{and} \quad 
   \kappa_{s_0} = \kappa_{\phi_0} = {1}/{R_0},
\label{eq:undef2}
\end{equation}
respectively.

All quantities introduced so far in this section carry the index ``$0$''
because they refer to the undeformed capsule configuration. The midsurface of
the deformed configuration is parameterized using analogous quantities without
indices ``$0$''. Specifically, its shape is determined by the sought-after
functions $r(s)$, $z(s)$ and the redundant $\psi(s)$. These functions have to
satisfy the boundary conditions $r(0) = r(L) = z(0) = 0$, $\psi(0) = 0$ and
$\psi(L) = \pi$, corresponding to a closed capsule without kinks at its poles.
The geometrical relations (\ref{eq:geom1}) and (\ref{eq:geom2}) can be
transferred directly to the deformed midsurface by omitting all indices
``$0$''.

\subsection{Measures of Deformation}

Having defined the parameterization of the deformed and undeformed midsurface,
we can now introduce measures of deformation. Fibers oriented along the
meridional and circumferential direction get stretched by the factors
\begin{equation}
 \lambda_s = \frac{\diff s}{\diff s_0} = s'(s_0) \quad \text{and} \quad 
  \lambda_\phi = \frac{r}{r_0} 
\label{eq:def_lambda},
\end{equation}
respectively. The function $s(s_0)$ defined in this context describes the
position $s$ at which a particle can be found that was originally located at
$s_0$. In order to center the measures of stretch around zero, we define the
\emph{meridional and circumferential strains} $e_s = \lambda_s -1$ and $e_\phi
= \lambda_\phi -1$. The bending of the midsurface is measured by the
\emph{meridional and circumferential bending strains}
\begin{equation}
 K_s= \lambda_s \, \kappa_s - \kappa_{s_0} \quad \text{and} \quad 
 K_\phi = \lambda_\phi \, \kappa_\phi - \kappa_{\phi_0}.
\end{equation}
They are more suitable than a simple difference of deformed and undeformed
curvature because they lead to more simple constitutive equations, as we will
see below.

\subsection{Elastic Law}

The strains measured by $e_s$, $e_\phi$, $K_s$ and $K_\phi$ give rise to
elastic tensions and bending moments in the capsule membrane. Assuming that
the capsule membrane consists of an hyperelastic material, there exists a
\emph{surface energy density} $w_S(e_s,\, e_\phi,\, K_s,\, K_\phi)$, which
measures the elastic energy that is stored in an infinitesimal patch of the
membrane divided by the area that this patch takes in the \emph{undeformed}
configuration. In this paper, we will use a Hookean model \cite{Libai1998}
\begin{multline}
 w_S =  \frac{1}{2} \frac{E\,H_0}{1-\nu^2} 
  \left( e_s^2 + 2\, \nu \,e_s\, e_\phi + e_\phi^2 \right) \\
   + \frac{1}{2} E_B  \left(  K_s^2 + 2 \,\nu\, K_s \,K_\phi + K_\phi^2
   \right)
\label{elastic_energy}
\end{multline}
with a (three-dimensional) Young modulus $E$, a bending modulus $E_B$, and a
Poisson ratio $\nu$, for a shell of (homogeneous) thickness $H_0$.  
Note that
the Poisson ratio of a two-dimensional membrane can take values $-1\leq \nu
\leq 1$, with $\nu=1$ corresponding to an area-incompressible membrane.
The
product $E\,H_0$ is frequently called the surface Young modulus.  In classical
small strain theory of plates, the bending modulus can be expressed as
\begin{equation}
 E_B = \frac{E\,H_0^3}{12(1-\nu^2)}. 
\label{eq:EBplate}
\end{equation} 
The bending energy contribution in the elastic energy 
 (\ref{elastic_energy}) agrees to leading order in $e_s$ and $e_\phi$ 
(where $K_s=\kappa_s-\kappa_{s_0}$ and $K_\phi=\kappa_{\phi}- \kappa_{\phi_0}$) 
and for an incompressible membrane  with  $\nu=1$
 with the commonly used Helfrich bending energy 
 $(\kappa_s + \kappa_\phi - c_0)^2$ (see
e.g.\ \cite{Quilliet2008,Marmottant2011}) with a  spontaneous curvature
$c_0=\kappa_{s_0}+\kappa_{\phi_0}$. 
The Helfrich bending energy was originally
proposed for two-dimensional liquids like vesicles, 
which differ qualitatively
from the solid shells we are considering. 

It can be shown
\cite{Libai1998} that the \emph{meridional tension} $\tau_s$ and 
\emph{bending  moment} $m_s$ derive from the surface energy density via
\begin{align}
 \tau_s &= \frac{1}{\lambda_\phi} \frac{\del w_S}{\del e_s}
   = \frac{E H_0}{1-\nu^2} \, \frac{1}{\lambda_\phi} 
  \big( e_s + \nu\, e_\phi \big), \\
    m_s &= \frac{1}{\lambda_\phi} \frac{\del w_S}{\del K_s}
   = E_B \, \frac{1}{\lambda_\phi}\big( K_s + \nu\, K_\phi \big).
  \label{stress-strain_2D}
\end{align} 
Likewise, we obtain the \emph{circumferential tension} $\tau_\phi$ and
\emph{bending moment} $m_\phi$ by analogous formulae with indices $\phi$ and
$s$ interchanged. Tensions and bending moments are measured per unit length of
the \emph{deformed} capsule, which is the reason why prefactors
$1/\lambda_\phi$ appear in these constitutive equations.

\begin{figure}
  \centerline{\includegraphics[width=60mm]{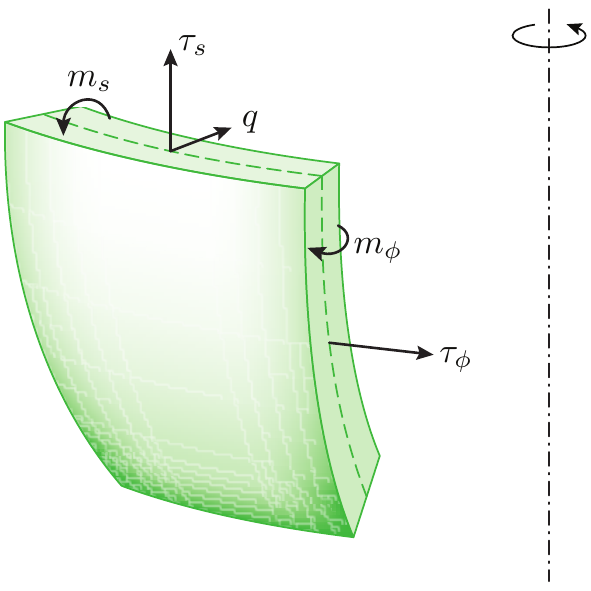}}
  \caption{(Color online) Tensions and bending moments acting on the faces
      of a membrane patch}
\label{fig:tensions}
\end{figure}

Figure \ref{fig:tensions} shows on which faces of a membrane patch the
tensions and bending moments act. Figure \ref{fig:tensions} also shows an
additional \emph{transverse shear tension} $q$ which acts on the top (and
bottom) side of the patch. It is constitutively undefined in our model because
we did not incorporate deformations in which the capsule's cross section gets
sheared 
(i.e.\ fibers normal to the midsurface get rotated). 
However, the transverse shear
tension is necessary to achieve force and moment equilibrium. Note that there
does not act any transverse shear tension on the right side of the patch
because of the assumption of axisymmetric, torsionless deformation.

\subsection{Equilibrium Conditions}

Besides the geometric relations and constitutive equations presented so far,
conditions of force and moment equilibrium are needed to close the
problem. These are three differential equations for tangential and normal
force equilibrium and moment equilibrium, which take the form
\cite{Libai1998,Pozrikidis2003}
\begin{align}
 0 &= - \frac{\cos \psi}{r}\, \tau_\phi 
  + \frac{1}{r}\, \frac{\diff(r\,\tau_s)}{\diff s} 
   - \kappa_s\,q, \label{eq:equil1}\\
 0 &= -p + \kappa_\phi \, \tau_\phi 
   + \kappa_s \, \tau_s 
   + \frac{1}{r}\, \frac{\diff(r\,q)}{\diff s}, \label{eq:equil2}\\
 0 &=  \frac{\cos \psi}{r} \, m_\phi 
  - \frac{1}{r} \frac{\diff(r\,m_s)}{\diff s} - q,\label{eq:equil3}
\end{align}
in the absence of external tangential force and torque densities and for a
constant pressure $p$ inside the capsule.

\section{Principle of Stationary Energy}

\subsection{Variation of Energy Functionals}

Another approach to find stable configurations of capsules for fixed but
altered volume $V\neq V_0$ is to minimize the functional of stored
elastic energy $F= \int w_S \; \diff A_0$ by calculus of variations. The
constraint of fixed capsule volume $V$ is handled by introducing a Lagrange
multiplier $p$ and extremizing the enthalpy $G=F-p\,V$ instead. The principle
of stationary total energy \cite{Libai1998} states that the solutions of the
equilibrium conditions (\ref{eq:equil1}) to (\ref{eq:equil3}) with given
pressure $p$ render the enthalpy $G$ stationary.

In the case at hand, this can be verified by using the standard
procedure of calculus of variations. Using $\diff A_0 = 2\,\pi\,r_0\,\diff
s_0$ for the area element of the undeformed midsurface and $V = \int
\pi\,r^2\;\diff z = \int \pi\,r^2\,\sin\psi\;\diff s = \int
\pi\,r^2\,\lambda_s\,\sin\psi\;\diff s_0$ as an integral expression for the
capsule volume, we arrive at the enthalpy
\begin{equation}
 G = \int_0^{L_0} \left( 2 \pi\, r_0\,w_S  
  - p \, \pi \, r^2\,\lambda_s\,\sin\psi\right) \; \diff s_0, 
 \label{eq:sphere_functional}
\end{equation} 
for which we have to find stationary points. 
The Euler-Lagrange equations are derived in
appendix \ref{appendix_variation} and turn out to be
\begin{eqnarray}
 0 &=& \frac{\cos \psi}{r} \, m_\phi 
 - \frac{1}{r}\,\frac{\diff(r\,m_s)}{\diff s} - q 
  \label{eq:EL_1}\\
 0 &=& \frac{1}{r}\,\frac{\diff(r\,\tau_s)}{\diff s} 
   - \frac{\cos \psi}{r} \, \tau_\phi 
  - \kappa_s \, q \label{eq:EL_2} \\
q &=& -\tau_s \, \tan \psi 
  + \frac{1}{2}\,p\,\frac{r}{\cos\psi}. 
\label{eq:sphere_q}
\end{eqnarray} 
They coincide with the general equations of tangential force and moment
equilibrium (\ref{eq:equil1}) and (\ref{eq:equil3}), but the normal force
equilibrium equation (\ref{eq:equil2}) is replaced by the algebraic expression
(\ref{eq:sphere_q}) for $q$. Instead of three differential equations in the
general case, we now end up with only two differential equations and one
algebraic relation. This means we have found a \emph{first integral} of the
general equations of equilibrium (which is only valid for capsules with
spherical resting shape under uniform pressure). Inserting the algebraic
expression (\ref{eq:sphere_q}) into the differential equation
(\ref{eq:equil2}) for $q$ confirms that the solution found here satisfies the
general equations of force equilibrium.

\subsection{Shapes with Opposite Sides in Contact}
\label{sec:contact}

\begin{figure}[t]
 \centering
 \includegraphics[width=65mm]{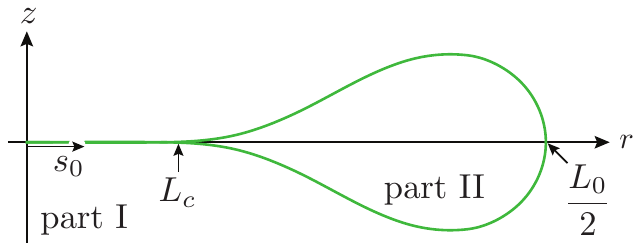}
 \caption{(Color online) Ansatz for the configuration with opposite sides in
   contact. Along the contour, the important coordinates $s_0 = L_c$ and $s_0
   = L_0/2$ are marked.}
 \label{fig:area_contact}
\end{figure}

Among the solutions for negative pressure, unphysical solutions that exhibit
self-intersection can occur. For two-dimensional shells with circular resting
shape (cylinder, ring), Flaherty {\it et\ al.}\ addressed this problem in
\cite{Flaherty1972}.  They found that the opposite sides of an elastic ring
touch at one point when the negative pressure reaches a certain threshold
$p_{c1}$.  Lowering the pressure further, the curvature at the point of
contact decreases until it finally becomes zero at a second critical pressure
$p_{c2}$. For pressures lower than $p_{c2}$, the contact area is a
straight-line segment, which increases in length with decreasing pressure.

In order to generalize these results to three-dimensional spherical shells, we
make the ansatz of a top/down symmetric deformed configuration with flat
circular areas around the poles in contact with each other, see figure
\ref{fig:area_contact}. Considering only top/down symmetric solutions, 
it is sufficient to set up the energy functional for the lower
hemisphere only. The flat part with $0 \le s_0 \le L_c$, is referred to as
part I, and the remaining part with $L_c \le s_0 \le {L_0}/{2}$ as part II. In
order to treat variations with respect to the boundary $L_c$ between parts I
and II properly, the enthalpy functional is split up into two corresponding
parts
\begin{align}
  G &= W-p\,V - \mu \, \Delta z \nonumber \\
    &= {\int_0^{L_c} \diff s_0 \big( 2\pi\,r_0 \, w_S \big)} \\
    &+ {\int_{L_c}^{L_0/2} \hspace{-10pt} \diff s_0\left( 2\pi \, r_0 \ w_S 
    - p \,\pi\,r^2\,\sin\psi\,\lambda_s 
      - \mu\,\lambda_s\,\sin\psi \right)}, \nonumber
  \label{eq:sphere_area_G}
\end{align} 
where the volume constraint and the condition $0=\Delta z = \int \diff z =
\int \lambda_s\,\sin\psi \; \diff s_0 $ are already incorporated via the
Lagrange multipliers $p$ and $\mu$.

Performing the first variation involves the following problems: In part I, the
function $r(s_0)$ must be varied, and in part II the functions $r(s_0)$ and
$\psi(s_0)$. Additionally, the integral boundary $L_c$ must be varied, which
yields continuity conditions at $s_0 = L_c$. The results can be summarized as
follows.

In part I, we obtain one Euler-Lagrange equation
\begin{equation}
 \frac{\del(r\,\tau_s)}{\del s} = \tau_\phi \qquad \text{for} \quad 
  s_0\in[0,\,L_c]. \label{eq:sphere_shape2}
\end{equation}
In part II we obtain a system of two Euler-Lagrange equations
\begin{align}
 0 &= \frac{\cos \psi}{r} \, m_\phi 
  - \frac{1}{r}\,\frac{\del(r\,m_s)}{\del s} - q 
  \label{eq:sphere_equil_area_1}\\
 0 &= \frac{1}{r}\,\frac{\del(r\,\tau_s)}{\del s} 
  - \frac{\cos \psi}{r} \, \tau_\phi - \kappa_s \, q 
  \label{eq:sphere_equil_area_2}\\
\quad q &= -\tau_s \, \tan \psi + \frac{1}{2}\,p\,\frac{r}{\cos\psi} 
  + \frac{\mu}{2\pi}\, \frac{1}{r\,\cos \psi} 
  \label{eq:sphere_area_q}
\end{align}
for $s_0 \in [L_c,\,L_0/2]$. The Euler-Lagrange equations for part II are very
similar to those obtained above for non-intersecting shapes, except for the
term $\frac{\mu}{2\pi}\, \frac{1}{r\,\cos \psi}$ in the algebraic expression
for $q(s_0)$. Nevertheless, it can be easily shown that relation
(\ref{eq:sphere_area_q}) also satisfies the general equation
(\ref{eq:equil2}), although it differs from the previous solution by the
additional term with the Lagrange multiplier $\mu$.

In order to fit the solutions of part I and II together, continuity conditions
must be derived, which arise from the variation with respect to the integral
boundary $L_c$. It turns out that most quantities of interest are continuous
at $s_0 = L_c$, namely $r$, $z$, $\psi$, $\lambda_i$, $\kappa_i$, $\tau_i$,
$m_i$ where an index $i$ stands for $s$ or $\phi$. Only for the transverse
shear tension $q$ no continuity condition derives because of the lack of a
constitutive relation between the shear tension $q$ and elastic deformations
in the present formulation. 
 We just know that the value $q(L_c)$ depends via the algebraic relation
(\ref{eq:sphere_area_q}) on the Lagrange multiplier $\mu$.

\section{Shape Equations}

The shape of the deformed capsule is governed by the equilibrium conditions,
constitutive equations and geometric relations presented above. They can be
rearranged to form a system of first order differential equations, known as
the \emph{shape equations},
\begin{equation}
 \begin{aligned}
  r'(s_0) &= \lambda_s \, \cos \psi \\
  z'(s_0) &= \lambda_s \, \sin \psi \\
  \psi'(s_0) &= \lambda_s \, \kappa_s \\
  \tau_s'(s_0) &= \lambda_s  
  \left( \cos \psi \, \frac{\tau_\phi - \tau_s}{r} + \kappa_s \, q \right) \\
  m_s'(s_0) &= \lambda_s 
  \left( \cos \psi \, \frac{m_\phi - m_s}{r} - q \right) \\
  q'(s_0) &= \lambda_s 
  \left( -\kappa_s \, \tau_s - \kappa_\phi \, \tau_\phi 
    - \cos \psi \,\frac{q}{r} + p\right).
 \end{aligned} 
 \label{eq:shape_hoo_shell}
\end{equation}
The first three equations in this system follow from the geometric relations
(\ref{eq:geom1}) and (\ref{eq:geom2}) (without index ``$0$'') and the last
three from the equilibrium conditions (\ref{eq:equil1}) to
(\ref{eq:equil3}). The change of variables from $s$ to $s_0$ was accomplished
by the relation $\diff s = \lambda_s \, \diff s_0$, see
(\ref{eq:def_lambda}). Usage of the algebraic expression (\ref{eq:sphere_q})
would reduce this system by one equation, but turns out to be numerically
impractical because of singularities when $\psi$ approaches $\pi/2$.

In order to close the above system of shape equations, all functions appearing
on the right hand side must be expressed in terms of the basic functions $r$,
$z$, $\psi$, $\tau_s$, $m_s$ and $q$. Exploiting geometrical relations, the
definitions of the strains and the constitutive equations, we find the set of
relations
\begin{align}
\begin{aligned}
 \kappa_\phi &= \frac{\sin \psi}{ r}, & 
 \lambda_s &= (1-\nu^2) \, \lambda_\phi \,  
   \frac{\tau_s}{E\,H_0} - \nu (\lambda_\phi -1) +1,\\
 \lambda_\phi &= \frac{ r}{ r_0}, & 
 \tau_\phi &= \frac{E\,H_0}{1-\nu^2} \, 
  \frac{1}{\lambda_s} \big( (\lambda_\phi-1) + \nu\, (\lambda_s -1) \big), 
\end{aligned} \nonumber \\
\begin{aligned}
  K_\phi &= \frac{\sin \psi - \sin \psi_0}{ r_0}, &
  K_s &= \frac{1}{ E_B} \, \lambda_\phi\, m_s - \nu\, K_\phi, \\
  m_\phi &=   E_B \, \frac{1}{\lambda_s}\big(  K_\phi + \nu\,  K_s \big),&
  \kappa_s &= \frac{ K_s+ \kappa_{s_0}}{\lambda_s},
\end{aligned}
\label{eq:shape_hoo_shell2}
\end{align} 
which close the system (\ref{eq:shape_hoo_shell}).

After changing variables from $s$ to $s_0$, the boundary conditions for a
closed capsule without kinks at its poles are $r(0) = r(L_0) = z(0) =\psi(0) = 0$ and $\psi(L_0) = \pi$, while the spherical reference shape
satisfies (\ref{eq:undef1}) and (\ref{eq:undef2}). These boundary conditions
cannot be directly used to determine boundary values for all strains and
curvatures using (\ref{eq:shape_hoo_shell2}) because some of the resulting
expressions are ill-defined as ratio of two vanishing quantities. These
boundary values can be evaluated analytically using L'H\^ospital's rule and
symmetry arguments (physically significant functions should be either odd or
even along the extended contour $s_0 \in [-L_0,\,L_0]$) in the limits $s_0
\rightarrow 0$ and $s_0 \rightarrow L_0$. Using this procedure 
boundary values at $s_0 = 0$ and $s_0 = L_0$ can be written as
\begin{equation}
 \begin{aligned}
 \lambda_s &= \lambda_\phi = \frac{E\,H_0}{E\,H_0 - \tau_s(1-\nu)}, 
   &  q' &= \lambda_s \left( \frac{p}{2} -\kappa_s\,\tau_s \right),
  \\
  \kappa_s &=  \kappa_\phi = \frac{ m_s}{E_B\,(1+\nu)} + \frac{1}{R_0 \,
    \lambda_s}, 
  & \tau_s' &=  m_s' = 0.
\end{aligned} 
\end{equation}

Now, the system (\ref{eq:shape_hoo_shell}) can be solved numerically. In order
to introduce dimensionless quantities, we choose $R_0$ as the length unit and
$E\,H_0$ as the tension unit. A simple shooting method and a multiple shooting
method \cite{numrec,stoer}, both with parameter tracing, were used to compute
capsule shapes for progressively lowered pressure $p<0$.

For top/down symmetric configurations it is sufficient to integrate from the
south pole $s_0 = 0$ to the equator $s_0 = L_0/2$. In this case, the boundary
conditions $r(0) = 0$, $z(0)=0$, $\psi(0)=0$, $q(0)=0$, $\psi(L_0/2)=\pi/2$
and $q(L_0/2)=0$
must be satisfied. Whereas the conditions concerning $r$, $z$ and $\psi$ are
obvious from geometry, the condition $q(0) = 0$ holds because of the algebraic
relation (\ref{eq:sphere_q}), and $q(L_0/2) = 0$ because of the symmetry of
the configuration: If $q(L_0/2) \neq 0$, the lower hemisphere would pull the
upper one to the inside (or outside, depending on the sign of $q(L_0/2)$),
which contradicts the top/down symmetry.

For solutions that are not top/down symmetric, we use shooting to a fitting
point \cite{numrec} because it is numerically impractical to integrate into a
removable singularity. The appropriate boundary conditions are $r(0) = 0$,
$z(0)=0$, $\psi(0)=0$, $q(0)=0$, $r(L_0)=0$, $\psi(L_0)=\pi$ and $q(L_0)=0$
in this case. The apparent problem that there are \emph{seven} boundary
conditions to a system of \emph{six} first order differential equations is
resolved by the algebraic relation (\ref{eq:sphere_q}). It effectively renders
one of the boundary conditions to $q$ obsolete (in other words, the
differential equation for $q$ with both boundary conditions concerning $q$ is
obsolete, which leads to a system of five differential equations with five
boundary conditions).

In the case of top/down symmetric configurations with opposite sides in
contact, the solutions of part I and II must be fitted together according to
the continuity conditions. In part I, the only degree of freedom is the
function $r(s_0)$ since $z(s_0) = \psi(s_0) = 0$ because of the geometrical
restrictions. The shape equations for part I can be obtained from the
equilibrium condition (\ref{eq:sphere_shape2}), together with constitutive and
geometric relations,
\begin{equation}
 \begin{aligned}
   r'( s_0) =& \lambda_s \\
  \lambda_s'( s_0) =& \frac{(1+\nu)}{R_0}\; 
  \frac{\cos(\frac{s_0}{R_0}) - 1}{\sin(\frac{s_0}{R_0})} 
  + \frac{r}{R_0^2\,\sin^2(\frac{s_0}{R_0})} \\
  &- \frac{\lambda_s \, \cot(\frac{s_0}{R_0})}{R_0}.
 \end{aligned} \label{eq:shape_sphere_reduced_2}
\end{equation} 
For reasons of symmetry, $\lambda_s(s_0)$ must be an even function along the
extended contour $s_0 \in (-L_0,\,L_0)$. Therefore, $\lambda_s'(0) = 0$. The
boundary condition for this part of the shape equations is $r(0)=0$. The
stretch at the pole, $\lambda_s(0)$, is free and serves as a shooting
parameter.

After a numerical solution of (\ref{eq:shape_sphere_reduced_2}) has been
found, the Hookean constitutive equations can be used to calculate the tension
$\tau_c \equiv \tau_s(L_c)$ and bending moment $m_c \equiv m_s(L_c)$, which
serve together with $r_c \equiv r(L_c)$ as initial values for the integration
on part II. The shape equations for part II can be adopted from
(\ref{eq:shape_hoo_shell}). Only the boundary conditions must be changed to
$r(L_c) = r_c$, $z(L_c)=0$, $\psi(L_c)=0$, $\tau_s(L_c)=\tau_c$, $m_s(L_c) =
m_c$, $z(L_0/2)=0$, $\psi(L_0/2)=\pi/2$ and $q(L_0/2)=0$. The shooting
parameters to satisfy the three conditions at the far end are $\lambda_s(0)$,
$q(L_c)$ and $L_c$.

In the case of top/down asymmetric configurations with opposite sides in
contact, we assume the region in contact to be a spherical cap with radius
$R$. A region $0\leq s_0 \leq L_c$ around the south pole has to change its
curvature from $1/R_0$ to $-1/R$ (note the sign change) to match a region $L_0
- L_{c2} \leq s_0 \leq L_0$ around the north pole. We allow the region around
the south pole to contract in the meridional direction by the factor
$\lambda_s(s_0) = \lambda_d$ (constant on $0\leq s_0 \leq L_c$). The region
around the north pole is allowed to contract by the factor $\lambda_s(s_0) =
\lambda_t$ (constant on $L_0 - L_{c2} \leq s_0 \leq L_0$) which may be
different from $\lambda_d$. Since the deformed configurations of these two
regions have to match exactly, we have the constraint $L_{c2} = L_c \,
\lambda_d / \lambda_t$.

Thus, the shape of the contact region of the deformed capsule is determined by
the four parameters $R$, $L_c$, $\lambda_b$ and $\lambda_t$. The shape of the
non-contacting part of the capsule is again described by the shape equations
(\ref{eq:shape_hoo_shell}). Since it is a system of six equations, and we have
four additional parameters that are not known a priori, we are able to satisfy
ten continuity conditions in total. That is just enough for the most important
quantities $r$, $z$, $\psi$, $\tau_s$ and $m_s$ at both ends $L_c$ and
$L_0-L_{c2}$ of the integration interval.

The explicit boundary conditions for the shape equations
(\ref{eq:shape_hoo_shell}) are given by
\begin{equation} 
\begin{aligned}
\textstyle r(L_c) &= \textstyle R\,\sin(L_c\,\lambda_d /R) \\
\textstyle z(L_c) &= \textstyle R\,(\cos(L_c\,\lambda_d /R) -1) \\
\textstyle \psi(L_c) &= \textstyle -L_c \,\lambda_d/R \\
\textstyle \tau(L_c) &= \textstyle \frac{EHo}{1-\nu^2} \frac{1}{\lambda_\phi} (\lambda_d -1 + \nu (\lambda_\phi - 1) ) \\
\textstyle m_s(L_c) &= \textstyle E_B\frac{1}{\lambda_\phi}(-\lambda_d / R - 1/R_0 + \nu (-\lambda_\phi/R-1/R_0)) \\
\textstyle \text{with} \quad \lambda_\phi &= \textstyle R\,\sin(L_c\,\lambda_d/R)/(R_0\,\sin(L_c/R_0))
\end{aligned}
\end{equation}
for the starting point and 
\begin{equation} 
\begin{aligned}
\textstyle r(L_0-L_{c2}) &= \textstyle R\,\sin(L_c\,\lambda_d /R ) \\
\textstyle z(L_0-L_{c2}) &= \textstyle R\,(\cos(L_c\,\lambda_d /R) -1) \\
\textstyle \psi(L_0-L_{c2}) &= \textstyle -L_c \,\lambda_d/R + \pi \\
\textstyle \tau(L_0-L_{c2}) &= \textstyle \frac{EHo}{1-\nu^2}\frac{1}{\lambda_\phi} ( \lambda_t -1 + \nu (\lambda_\phi - 1) ) \\
\textstyle m_s(L_0-L_{c2})&= \textstyle E_B \frac{1}{\lambda_\phi}(\lambda_t / R - 1/R_0 + \nu(\lambda_\phi/R-1/R_0)) \\
\textstyle \text{with} \quad \lambda_\phi &= \textstyle R\,\sin(L_{c2}\,\lambda_t/R)/(R_0\,\sin(L_{c2}/R_0))
\end{aligned}
\end{equation}
for the end point of integration.

Note that taking $\lambda_s$ to be constant in the regions in contact is a
strong simplification. A more involved theory would incorporate equations
similar to (\ref{eq:shape_sphere_reduced_2}) to determine the
in-plane-displacements. However, we observed that the solution branches
produced with this simplified method fit neatly in the bifurcation diagrams.

\section{Bifurcation Behavior}

As the surface Young modulus $E\,H_0$ serves as the tension unit, there are
only two elastic parameters left to vary, the Poisson ratio $\nu$ and the
dimensionless bending modulus
\begin{equation}
\tilde E_B \equiv  \frac{E_B}{R_0^2 \, E\,H_0} = \frac{ H_0^2}{12(1-\nu^2)R_0^2}.
\label{eq:tildeEB}
\end{equation}
where we used (\ref{eq:EBplate}) for the bending modulus of elastic plates. 
We will present bifurcation diagrams for a Poisson ratio of $\nu=0.5$ and
dimensionless bending moduli of $\tilde E_B = 0.01$ and $0.001$. 
These values
correspond to relative shell thicknesses of $H_0 / R_0 = 0.3$ and $0.095$,
respectively.

\subsection{Spherical Solution Branch}

The trivial branch of spherical solutions (branch 1 in the 
bifurcation diagrams below)
can be calculated analytically because of its high
symmetry. Solving the equilibrium equations
(\ref{eq:equil1}) to (\ref{eq:equil3}) for  a
sphere with radius $R$, it is straightforward to show that these equations 
reduce to the modified Laplace-Young equation 
\begin{equation}
  p = 2 \, \kappa \, \tau,
\end{equation}
 with $\kappa = 1/R$ the isotropic
curvature and $\tau \equiv \tau_s = \tau_\phi$ the isotropic tension. The
Laplace-Young equation
 determines the new radius of the capsule for given pressure. Using
the Hookean constitutive relation to express $\tau$ in terms of the isotropic
stretch $\lambda_s = \lambda_\phi = R/R_0$, this condition can be rewritten
as 
\begin{equation} 
R^2 - \frac{2\,EH_0}{p\,(1-\nu)} \, (R-R_0)  = 0.
\label{Rp1}
\end{equation}
The solution to this quadratic equation is the radius-pressure relation
of the spherical branch (called branch 1 below)
\begin{equation}
 R_{1} = \frac{EH_0}{p\,(1-\nu)} \pm 
 \sqrt{ \left(\frac{EH_0}{p\,(1-\nu)}\right)^2 
  - \frac{2\, EH_0}{p\, (1-\nu)} \, R_0 },
\label{Rp}
\end{equation}
where the $+$ branch 
 holds for $p<0$ and the $-$ branch for $p>0$, which can
easily be inferred from requiring  $\lim_{p\rightarrow 0} R =
R_0$. This radius-pressure relation determines the deformed configuration
completely, and all physical properties, like volume, strains, tensions and
stored elastic energy, can be calculated in turn.

In the following we focus on buckling shapes for negative pressure $p<0$. 
For positive pressure $p>0$, where capsules are stretched, 
the spherical branch 1 represents 
 the {\em only} equilibrium shape.

\subsection{Stability Criteria}

We have shown that the equations of force and moment equilibrium render the
functional of total energy stationary. When the shape equations are solved for
negative pressure, various solution branches with reduced capsule volume
$V<V_0$ can be found, which can represent local minima or maxima 
of the energy functional. 
We display the total energy or enthalpy of each solution branch in
an energy bifurcation diagram as a function of the capsule volume $V$ or the
pressure $p$, respectively. 
The principle of minimal total energy allows us to determine the
globally stable branch as the branch of minimal energy among all stationary
shapes. Shape transitions such as buckling occur where two branches intersect.

If the capsule volume is given, the stored elastic energy
\begin{equation}
 F = \int_0^{L_0} 2\pi\,r_0\,w_S\; \diff s_0
\end{equation}
must be minimal. This criterion has experimental significance if the capsule
volume cannot change because the encapsulated liquid is incompressible and the
membrane is impermeable. In cases of a semipermeable capsule membrane, it is
also reasonable to consider the volume fixed because the relaxation into the
equilibrium shape happens on much shorter time scales than the diffusion of
the inner liquid through the membrane. In the corresponding bifurcation
diagram we display the stored elastic energy $F$ as a function of the volume
$V$.

On the other hand, if the capsule is filled and surrounded by gases, the
pressure difference $p$ is prescribed rather than the capsule volume. In this
case, configurations with minimal enthalpy
\begin{equation}
 G = F - p\,V = \int_0^{L_0} \left( 2 \pi\, r_0\,w_S  
  - p \, \pi \, r^2\,\lambda_s\,\sin\psi\right) \; \diff s_0
\end{equation} 
are energetically preferable. In the corresponding bifurcation diagram we
display the elastic enthalpy $G$ as a function of the pressure $p$. Note that
$G(p)$ is the Legendre transform of $F(V)$, since the relation $p = \diff F /
\diff V$ holds (which was verified numerically for all solution branches
presented here). Solution branches that lie lowest in the $F(V)$ bifurcation
diagrams need not necessarily coincide with the lowest branches in the $G(p)$
diagrams.

Finally, it is also useful to analyze the relation between pressure $p$ and
volume $V$ for stationary capsule shapes. Branches that exhibit an unusual
pressure-volume relation with $\diff p / \diff V < 0$ are inherently unstable
if the pressure is given instead of the volume \cite{Balloons}. To see that,
we consider a water filled capsule connected to a reservoir of water. The
pressure in this system can be prescribed. Now, if we try to change the
capsule volume by lowering the pressure by an amount $\diff p < 0$, the
capsule grows by an amount $\diff V >0$, i.e.\ water flows from the reservoir
into the capsule. The loss of water in the reservoir typically leads to an
even lower pressure and thus, the equilibrium is unstable with repsect 
to volume changes.
This instability 
is also reflected by a negative 
second derivative of the free energy, 
$d^2F/dV^2 = \diff p / \diff V < 0$
and a horizontal $p(V)$ curve with 
 $dp/dV=0$ marks the onset of such an instability.

Therefore, we have three 
 criteria of stability for capsule shapes: 
\begin{itemize}
\item[(i)]
minimal energy $F$ for fixed capsule volume $V$, 
\item[(ii)]
 minimal enthalpy $G$ for
fixed pressure $p$, and 
\item[(iii)] a monotonously decreasing pressure-volume
relation $\diff p /\diff V < 0$ is sufficient
for an unstable shape for fixed pressure. Thus, 
the criterion  $\diff p /\diff V \ge 0$ is 
only a   {\em necessary} condition for stability:  
Configurations with $\diff p /\diff V \ge 0$ can still be unstable
with respect to deformation modes which do not change the volume.
\end{itemize}

In fact, criterion (iii) can be generalized by using a 
general bifurcation theorem that has been proven in 
Ref.\ \cite{Maddox87}. This allows us 
to make a further statement about the 
instability of shapes beyond points where a 
  monotonously decreasing $p(V)$ curve becomes vertical:
If the  $p(V)$ curve  at such a 
point is open to the left, i.e., the following lower part 
of the $p(V)$ curve has again a positive slope  
$dp/dV>0$, this lower part must also be {\em unstable}.
Furthermore, this is an instability 
with respect to a volume-preserving mode. 
This generalization also demonstrates
that a positive slope  $\diff p /\diff V \ge 0$ is 
not sufficient for stability.

\subsection{Bifurcation Diagrams}

 \begin{figure*}[p]
 \centering
 \includegraphics[width=160mm]{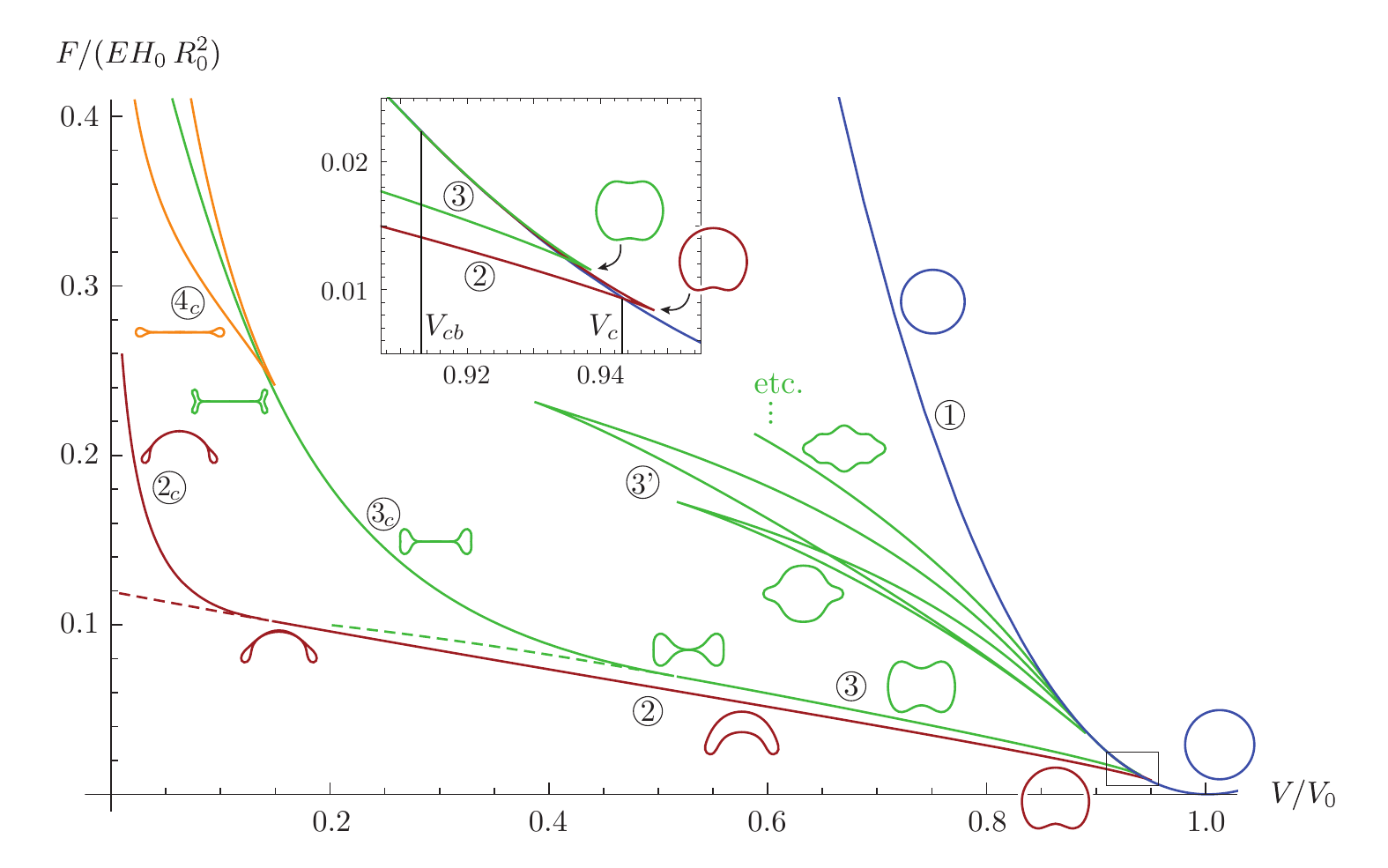}
 \caption{(Color online) Bifurcation diagram for given volume of a capsule
   with $\tilde E_B = 0.001$. On the dashed lines of branch 2 and 3, the
   capsule intersects itself. Branch 3' continues winding up in the diagram,
   as indicated by the "etc ...".}
 \label{fig:001_L1}
\end{figure*}

\begin{figure*}[p]
 \centering
 \includegraphics[width=160mm]{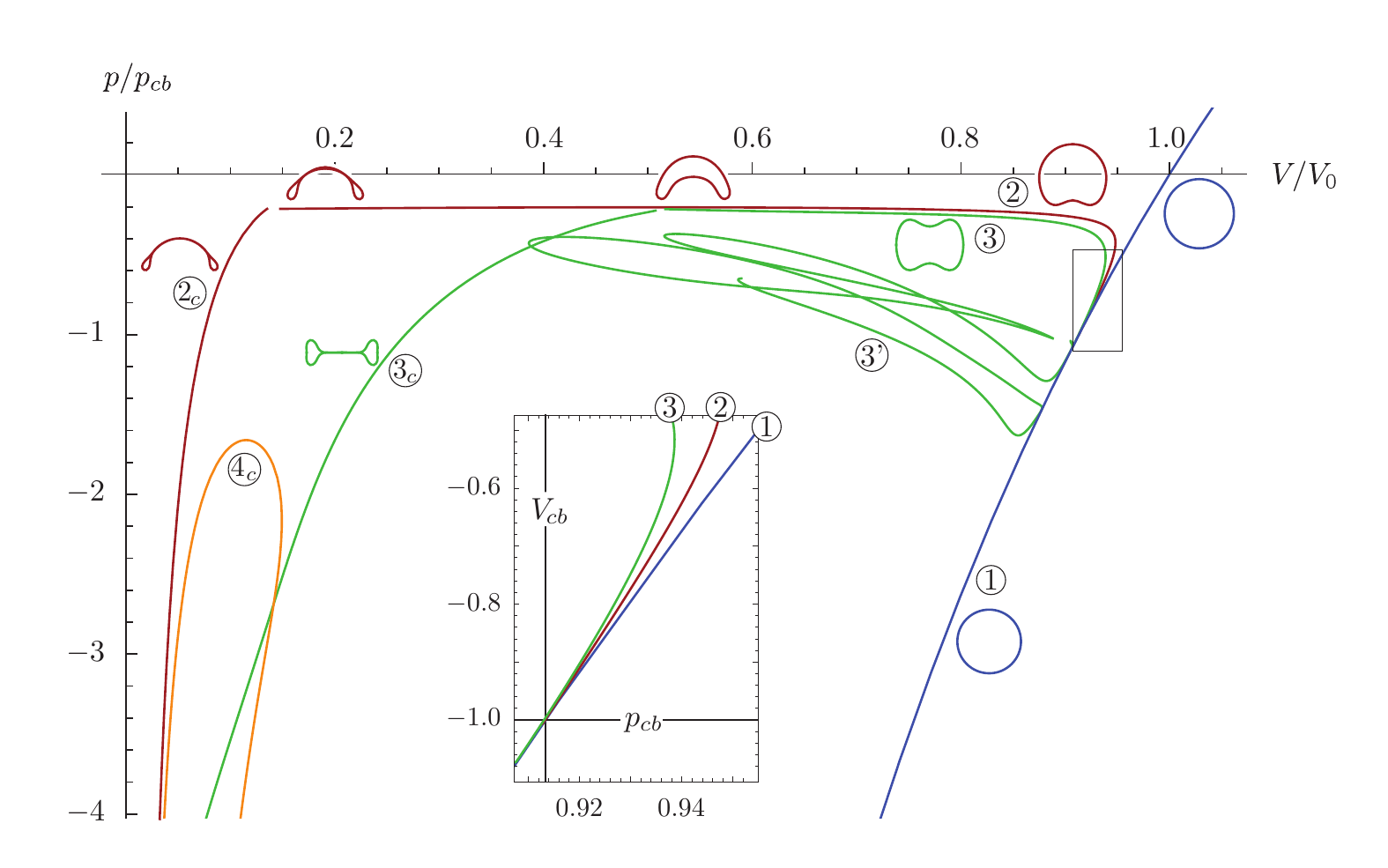}
 \caption{(Color online) Pressure-volume relation of a capsule with $\tilde
   E_B = 0.001$}
 \label{fig:001_L2}
\end{figure*}

\begin{figure*}[t]
 \centering
 \includegraphics[width=160mm]{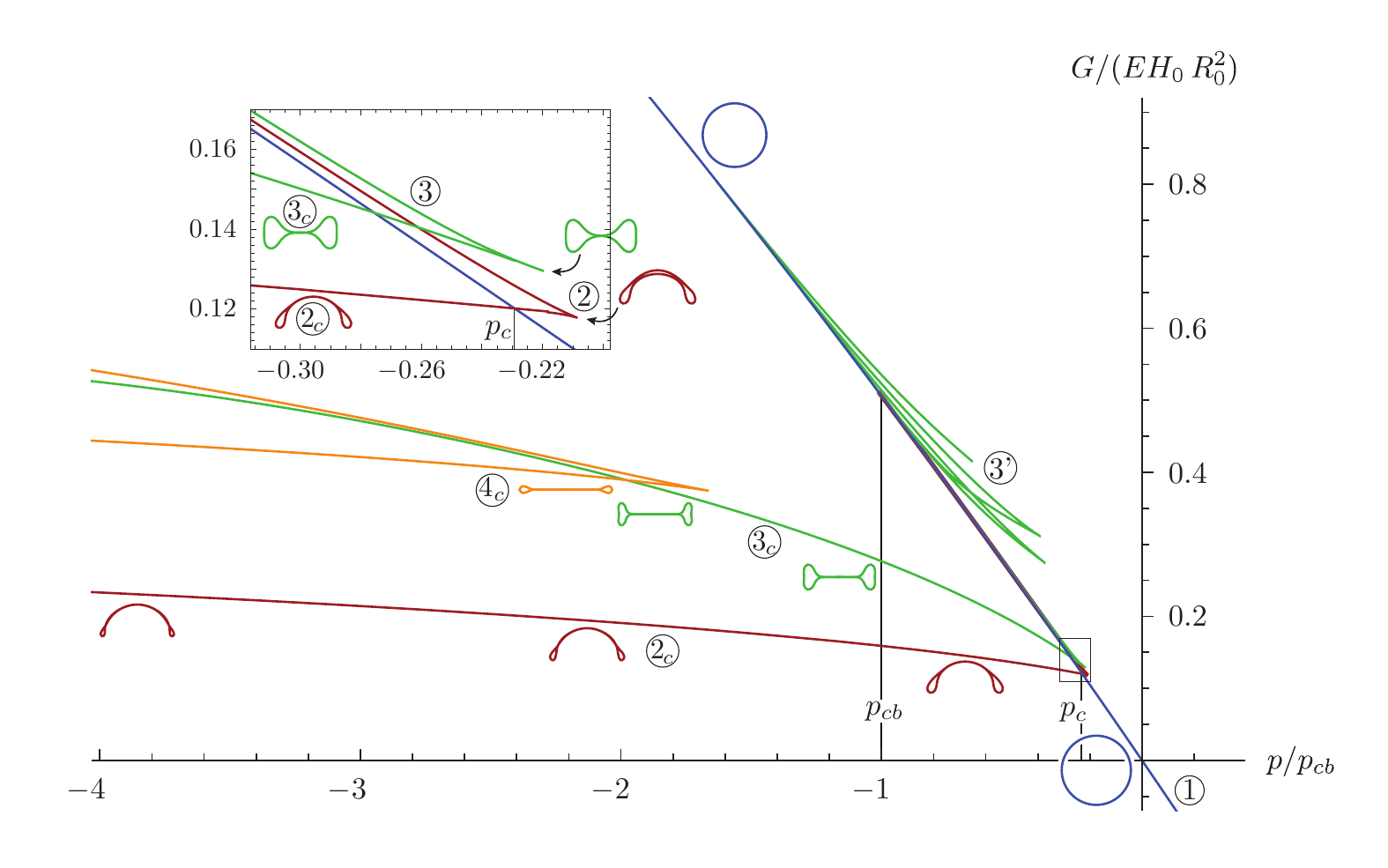}
 \caption{(Color online) Bifurcation diagram for given pressure of a capsule
   with $\tilde E_B = 0.001$.}
 \label{fig:001_L3}
\end{figure*}

\begin{table}
\begin{center}
\begin{tabular}{ccc}
  \hline
  Branch	&  Configuration  &  Example \\ \hline 
  1   		&  spherical & \raisebox{-6pt}{\includegraphics[height=18 pt]{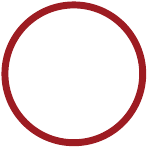}}  \\ 
  2   		&  simply buckled (asymmetric) & \raisebox{-6pt}{\includegraphics[height=18 pt]{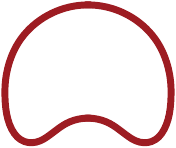}}\\ 
  2'  		&  asymmetrically crumpled & \raisebox{-6pt}{\includegraphics[height=18 pt]{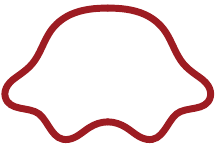}}\\ 
  $2_c$	&  simply buckled with contact & \raisebox{-6pt}{\includegraphics[height=18 pt]{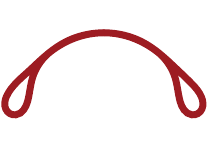}}\\
  3  		&  symmetrically buckled & \raisebox{-6pt}{\includegraphics[height=18 pt]{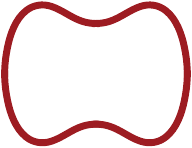}}\\ 
  3', 3'' 	&  symmetrically crumpled & 
      \raisebox{-6pt}{\includegraphics[height=18 pt]{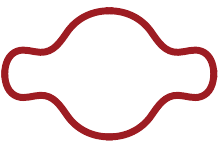}}, \raisebox{-6pt}{\includegraphics[height=18 pt]{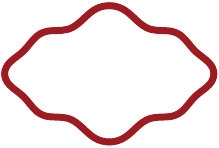}}\\ 
  $3_c$, $4_c$	&  symmetrically buckled with contact & 
    \raisebox{-6pt}{\includegraphics[height=18 pt]{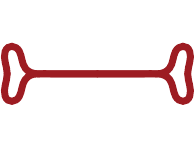}},  \raisebox{-6pt}{\includegraphics[height=18 pt]{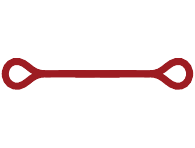}} \\
  \hline
\end{tabular}
\caption{\label{table:branches}
Denomination of different branches in bifurcation diagrams.
Branches are classified according to the number of bulges 
as  buckled (one or two bulges) or crumpled (more than two bulges, names with primes) 
and according to top/down mirror-symmetry (shapes 1,3,4 are top-down
symmetric, shapes 2 are asymmetric). 
Shapes with subscript $c$ exhibit 
contact of originally opposite sides; shape $3_c$ develops an additional dimple 
at the sides as opposed to shape $4_c$ (for $\tilde E_B = 0.001$).}
\end{center}
\end{table}

With these  criteria of stability, we analyze the bifurcation
behavior of elastic  capsules by investigating the 
different branches in three types of bifurcation diagrams: 1) In the 
$F(V)$ diagram we study buckling by reducing the capsule volume by 
using criterion (i).
2) In the $p(V)$ diagram we can identify unstable
shapes as  monotonously decreasing branches
 $\diff p /\diff V < 0$ according to criterion (iii).
3) In the $G(p)$ diagram the Legendre transform 
$G(p) = F(V(p)) -p V(p)$ allows us to  investigate buckling 
under negative pressure according to criterion (ii). 

The nomenclature of solution branches is summarized in table
\ref{table:branches}.  In particular, we will compare our results to classical
buckling theory \cite{Pogorelov1988,LL7,Libai1998}, which predicts a critical
negative buckling pressure
\begin{equation}
   p_{cb} = -4 \frac{E\,H_0^2}{R_0^2 \, \sqrt{12(1-\nu^2)}} 
 = -4 \sqrt{EH_0 \, E_B} / R_0^2
\label{pcb}
\end{equation}
for a spherical capsule. 
A corresponding critical volume can be obtained as 
\begin{equation}
   V_{cb} = \frac{4\pi}{3} \, (R_{1}(p_{cb}))^3
\end{equation}
by using $p_{cb}$ in the 
radius-pressure relation (\ref{Rp})  of the spherical branch 1.

We start with the bifurcation
behavior of thin capsules with $\tilde E_B = 0.001$, as shown in the diagrams
\ref{fig:001_L1} to \ref{fig:001_L3}. 
The $F(V)$ diagram (figure \ref{fig:001_L1}) reveals that the simply buckled
configurations of branch 2 are 
{\em energetically} favorable for volumes $V<V_c$, where
$V_c$ denotes a critical volume which is $V_c \approx 0.944\,V_0$ in this
case. At the critical volume  $V_c$, the spherical branch 1 
and branch 2  of  simply buckled configurations   intersect. 
Note that this volume is {\em larger}  than the critical volume 
$V_{cb}$ within classical buckling theory, which is 
$V_{cb} \approx 0.914 \,V_0$ for this case.
A shape transition between 
 branches 1 and 2 at $V_c$ is {\em discontinuous}
and involves an energy barrier.
The upper part of  branch 2 most likely represents 
 the unstable transition state 
at $V=V_c$ between a spherical shape 
 and the stable lower part of branch 2. Therefore, the energy barrier 
can be estimated by the energy difference between the upper 
and lower parts of branch 2  at  $V=V_c$. 
More detailed stability considerations are given in section \ref{sec:buckling}
below.

For volumes $V<0.14\,V_0$, these configurations start to intersect themselves
(dashed line).  Simultaneously, a branch $2_c$ starts to exist with simply
buckled configurations with opposite sides in contact. Although the method
used to obtain this branch incorporated some simplifications, branches $2$ and
$2_c$ connect neatly in the diagrams. In this domain, the ansatz with opposite
sides in contact produces solutions with higher energies (branch $2_c$
compared to the dashed part of $2$), i.e.\ the self-intersecting solution
branch raises in the $F(V)$ diagram when it is forced to satisfy the physical
constraints.

For volumes $V<V_c$, 
the energetically next best configurations according to (i)  are the
symmetrically buckled ones of branch 3. They exhibit self-intersection for
volumes $V<0.51 \, V_0$. At higher energies, crumpled
configurations can be found in branch 3' that is connected to branch 3. They
correspond to local minima, saddle points or maxima of the energy functional. 
Several more crumpled shapes
can be found in this domain of the bifurcation diagram, but they are not shown
for the sake of clarity.
Similar crumpled configurations have been observed for small 
volumes  in simulations using
triangulated  surface models, in particular at high 
compression rates \cite{Vliegenthart2011}.
Trapping in metastable crumpled shapes  can
 contribute to this behavior. 
 
 \begin{figure*}[p]
 \centering
 \includegraphics[width=160mm]{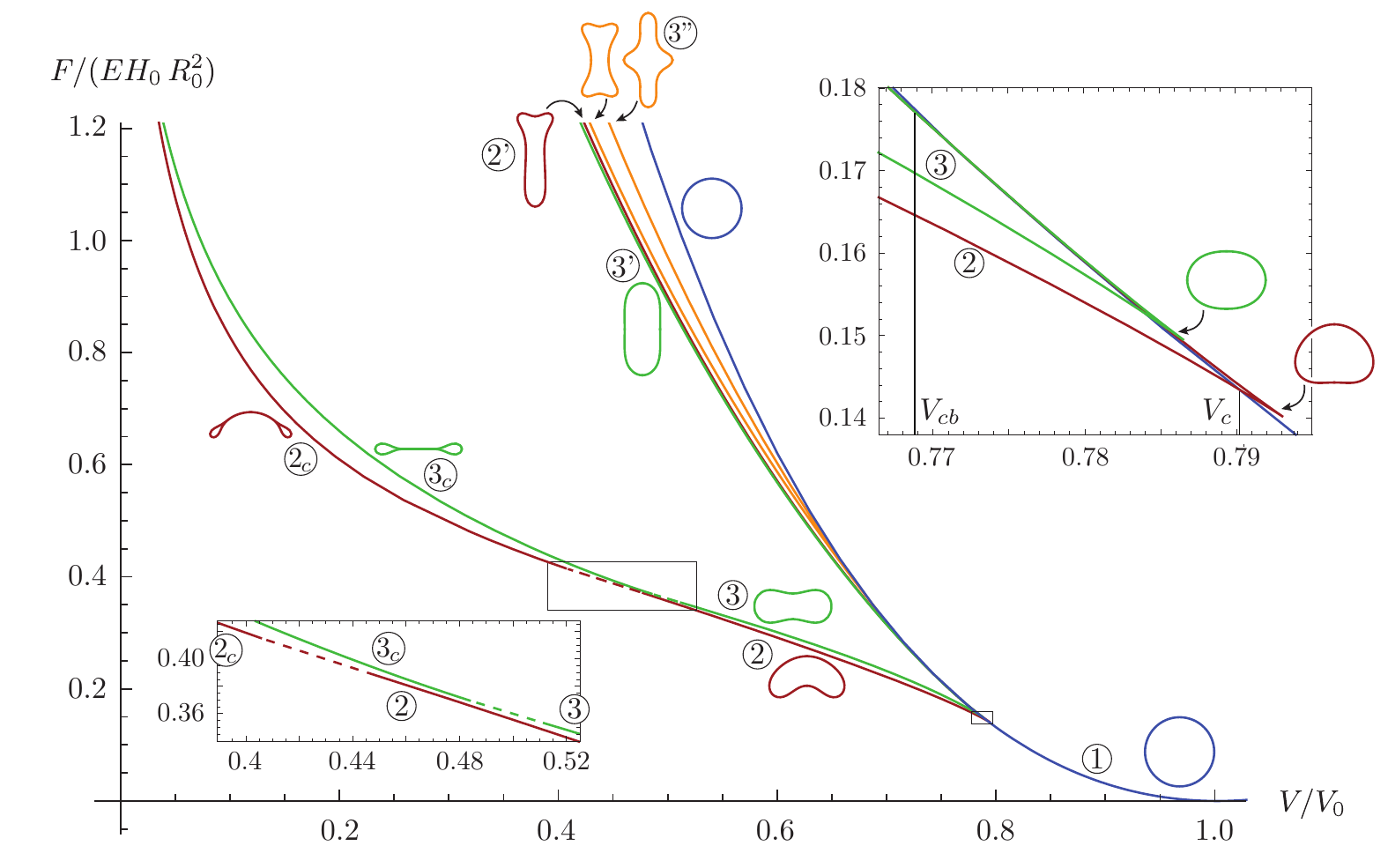}
 \caption{(Color online) Bifurcation diagram for given volume of a capsule
   with $\tilde E_B = 0.01$. The dashed lines in the gaps between $2/2_c$
   and $3/3_c$ could not be calculated numerically, but we expect some
   configurations to exist in these domains.}
 \label{fig:01_L1}
\end{figure*}

\begin{figure*}[p]
 \centering
 \includegraphics[width=160mm]{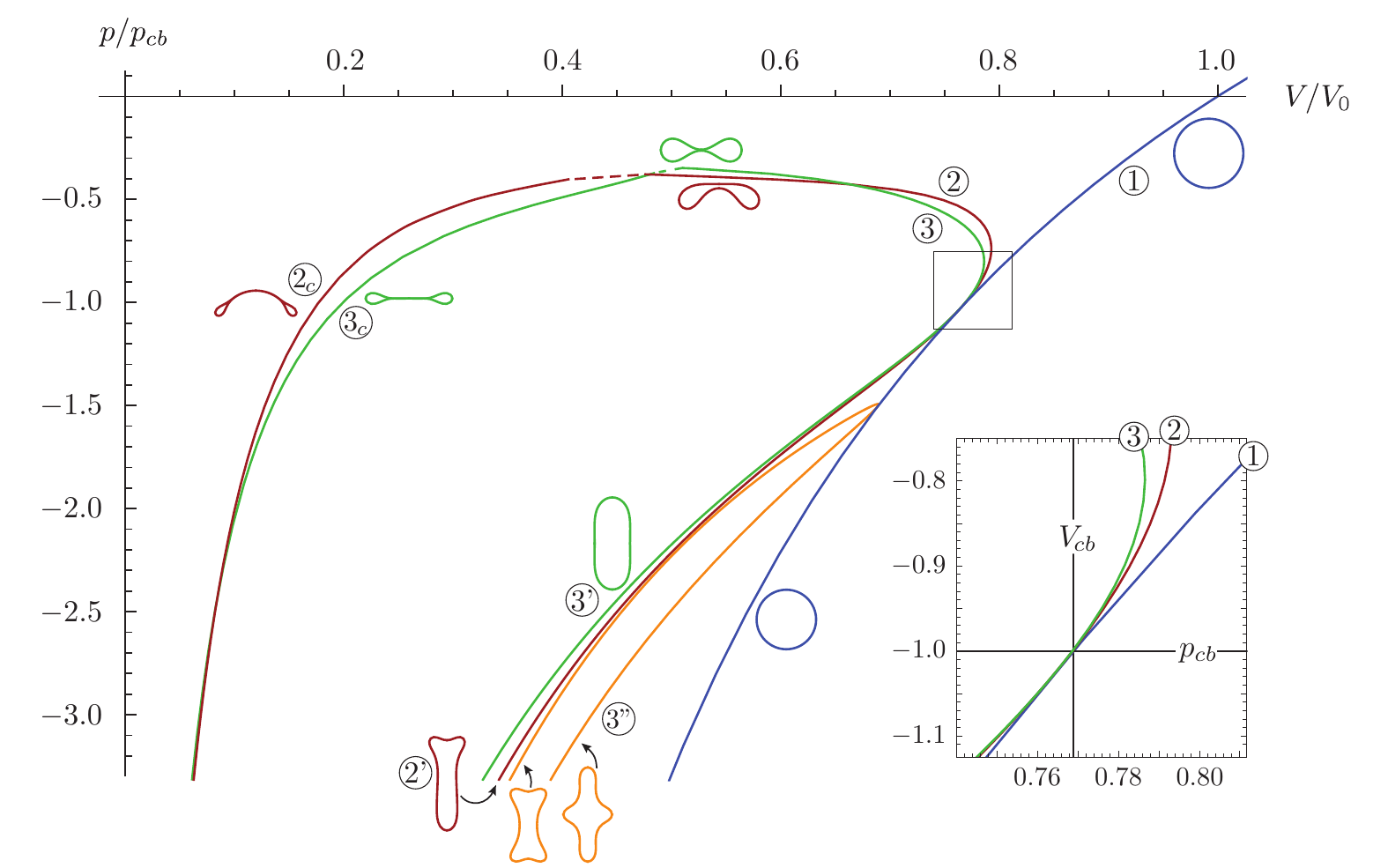}
 \caption{(Color online) Pressure-volume relation of a capsule with $\tilde
   E_B = 0.01$. The dashed lines in the gaps between $2/2_c$ and $3/3_c$
   could not be calculated numerically, but we expect some configurations to
   exist in these domains.}
 \label{fig:01_L2}
\end{figure*}

\begin{figure*}[t]
 \centering
 \includegraphics[width=160mm]{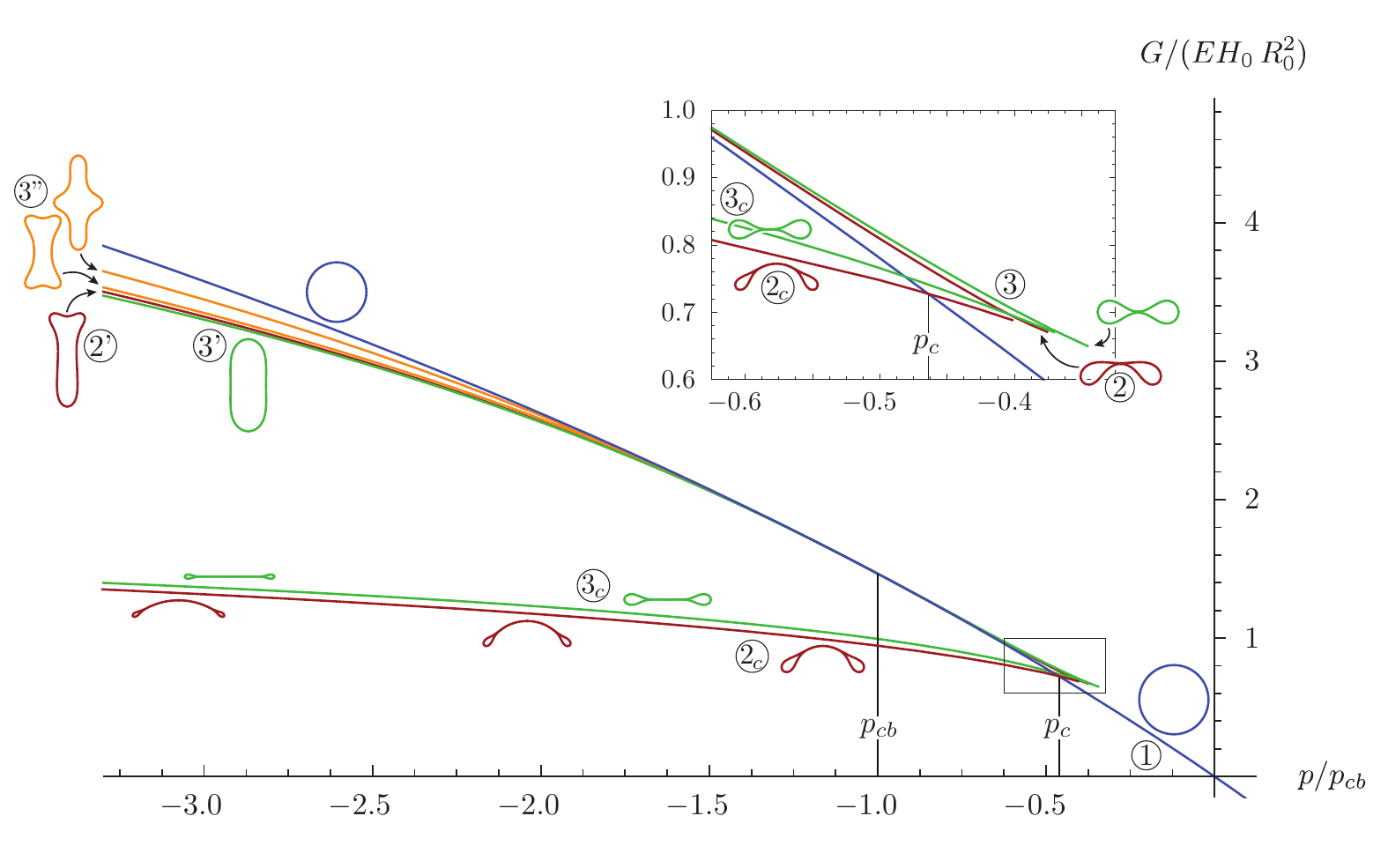}
 \caption{(Color online) Bifurcation diagram for given pressure of a capsule
   with $\tilde E_B = 0.01$}
 \label{fig:01_L3}
\end{figure*}

For given pressure, the $p(V)$ diagram (figure \ref{fig:001_L2}) and $G(p)$
diagram (figure \ref{fig:001_L3}) can be analyzed in order to identify
unstable  solutions according to criteria (ii) and (iii).
The simply and symmetrically buckled solutions of branches 2
and 3 (without contact of opposite sides) exhibit a negative slope in the
$p(V)$ diagram. 
This means that they are mechanically unstable with respect to volume 
reducing  deformations for given
pressure according to criterium (iii),
 although they are  most stable for fixed volume 
in the $F(V)$ diagram.
According to the generalization of criterion (iii) 
based on  Ref.\ \cite{Maddox87} also the lower part of branches 2
and 3, which lie  beyond the turning point and
 have a positive slope in   
the $p(V)$ diagram at volumes $V> V_{cb}$, 
(see also inset in figure \ref{fig:001_L2}) are unstable, 
however, with respect to volume-preserving modes.  
The energy diagram
$G(p)$ confirms this result; the unstable 
 branches 2 and 3 lie above the
trivial solution branch 1. On the other hand, the buckled
configurations $2_c$ and $3_c$ with opposite sides in contact are 
mechanically stable again,
with $\diff p / \diff V >0$, and $2_c$ is energetically preferable.

At 
the critical pressure  $p_c \approx - 0.23 \, |p_{cb}|$, 
the spherical branch 1 
and branch $2_c$ of simply buckled
configurations with opposite sides in contact  intersect
in the $G(p)$ diagram,
 and shapes $2_c$ become {\em energetically} favorable for given pressure.

The negative critical pressure $|p_c|$ is  much smaller  than the 
 classical buckling pressure $|p_{cb}|$, which is 
$p_{cb}\approx - 0.126 \, EH_0/R_0$  according to eq.\ (\ref{pcb}).
However, the classical buckling pressure fits very well to the point, where
branches  2 and 3 emerge from the spherical solution branch 
in the $p(V)$ diagram  (see inset in figure
\ref{fig:001_L2}, where it is indicated by a horizontal line). 
The classical buckling pressure $p_{cb}$ is the pressure where the 
unbuckled spherical configuration 1 becomes unstable with respect to 
buckling \cite{LL7,Ru2009}, and a spontaneous transition to the 
unstable simply buckled branch 2  occurs. 

At the critical pressure $p_c$, on the other hand, 
  shape $2_c$ becomes {\em energetically} favorable as compared 
to the spherical branch 1. 
A shape transition between 
both branches at $p_c$ is {\em discontinuous}
 and involves an energy barrier.
The upper unstable branch 2 of a simply buckled shape 
without contact most likely represents the transition state 
 between a spherical shape 
 and branch $2_c$ at $p=p_c$. Therefore, the energy barrier 
can be estimated by the energy difference between the upper 
unstable branch 2  and the lower stable  branch $2_c$  at  $p=p_c$.

The bifurcation diagrams for a thick capsule with $\tilde E_B = 0.01$ (figures
\ref{fig:01_L1} to \ref{fig:01_L3}) look qualitatively similar. Again, the
simply and symmetrically buckled solution branches 2 and 3, respectively, 
are energetically
preferable for given volume, but exhibit a negative slope in the $p(V)$
diagram (figure \ref{fig:01_L2}), and are unstable and 
energetically unfavorable for given
pressure.

For given volume $V$, 
the $F(V)$ diagram (figure \ref{fig:01_L1}) shows that the simply buckled
configurations of branch 2 are energetically lower than 
spherical shapes  for  $V<V_c$ with a 
critical volume  $V_c \approx 0.79 \,V_0$, which is 
again larger than the  critical volume $V_{cb} \approx 0.77 \,V_0$
from  classical buckling theory.

For given pressure $p$, 
the $G(p)$ diagram (figure \ref{fig:01_L3}) shows  that 
simply buckled
configurations $2_c$ with opposite sides in contact are 
energetically favorable as compared to spherical shapes of branch 1 
for $p<p_c$, where the critical pressure 
is  $p_c \approx -  0.46\, |p_{cb}|$.
Also for  $\tilde E_B = 0.01$, 
the negative critical pressure $|p_c|$ is  much smaller  than the 
 classical buckling pressure $|p_{cb}|$, which is 
$p_{cb}\approx - 0.4 \, EH_0/R_0$  in this case and 
 the classical buckling pressure fits to the point, where
branch 2 emerges from the trivial spherical 
solution branch (see inset in figure
\ref{fig:01_L2}, where it is indicated by a horizontal line). 

For $\tilde E_B = 0.01$, 
 there is a visible gap between the buckled branches 2 and 3
and their respective continuations $2_c$ and $3_c$ 
with opposite sides in contact. This gap was
already present in the diagrams for $\tilde E_B = 0.001$, but much smaller. It
is assumed to be closed by configurations with point contact of north and
south pole. In the case of branch 3, an analogous behavior like 
that of elastic rings is expected \cite{Flaherty1972}. 
The curvature at the point of contact is expected to decrease, until it
finally becomes zero and hence fulfills the continuity conditions for circular
areas in contact. However, the shape equations when point contact of north and
south pole is enforced turn out to be hard to solve numerically, because the
transverse shear tension diverges at the poles.

At higher energies, there are again configurations with several bulges
(branches 3' and 3''). In the $F(V)$ diagram (figure \ref{fig:01_L1}), they
lie lower than the trivial solution branch. In contrast to the results for the
capsule with $\tilde E_B = 0.001$, branch 3'' does not have 
 multiple turning points
and is not connected to the continuation 3' of the symmetrically buckled
branch within the scope of our diagrams. However, branches 3' and 3'' might
join at higher energies and lower pressures. Notably, these solution branches
lie lower than branch 1 in the $G(p)$ diagram (figure \ref{fig:01_L3}), which
is a qualitative difference to the results for $\tilde E_B = 0.001$.

The $F(V)$ and $p(V)$
diagrams are in good agreement with previous work based on
triangulated surface models: In Ref.\ \cite{Marmottant2011}, a
$p(V)$ relation has been obtained, which  also shows a
uniform shrinkage of the capsule (our branch 1) for small volume 
reduction followed by a 
jump into an axisymmetric simply buckled configuration (our branch 2)
with the same $p(V)$ behavior as branch 2.
Furthermore, Ref.\ \cite{Quilliet2008} contains an $F(V)$ diagram 
in which the
spherical and simply buckled solution branches are shown. They reveal
qualitatively the same $F(V)$ relation as our branches 1 and 2.

\subsection{Buckling Bifurcation} 
\label{sec:buckling}

\begin{figure*}
 \centering
 \includegraphics[width=172mm]{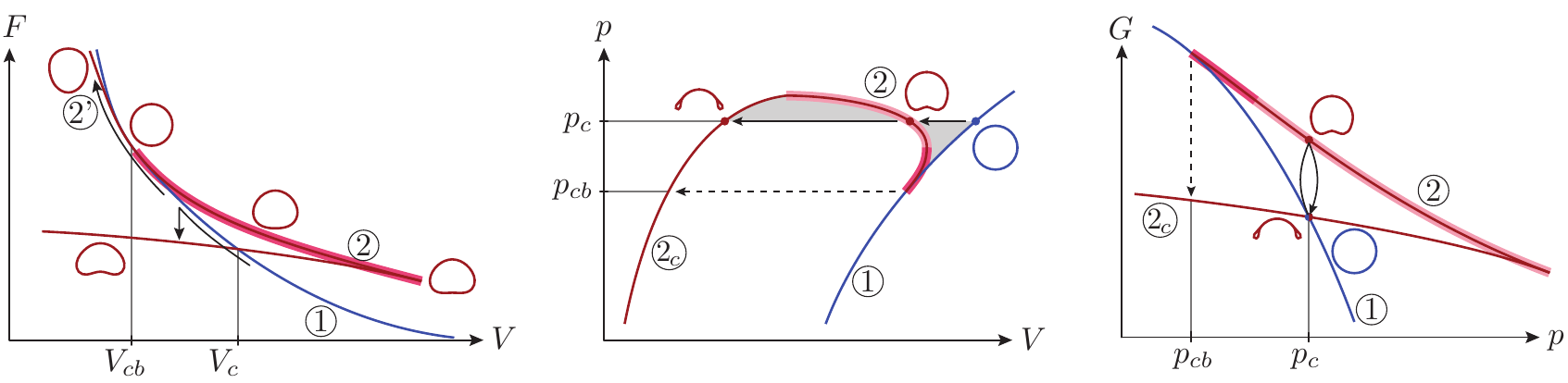}
 \caption{(Color online) Left: Schematic drawing of two different
   bifurcation phenomena (indicated by the arrows) for given volume $V < V_c$.
   Middle and right: Schematic drawing of a realistic buckling process for
   given pressure. The critical pressure $p_c$ is defined by the crossing of
   branches 1 and $2_c$ in the $G(p)$ diagram. As a consequence of the
   Legendre transformation, the two 
  regions shaded gray in the $p(V)$ diagram have
   the same area at $p=p_c$ (Maxwell construction).}
 \label{fig:bif}
\end{figure*}

The key features of the bifurcation diagrams presented above are drawn
schematically in figure \ref{fig:bif}. They allow to construct a 
complete picture of the bifurcation behavior at the 
buckling transition.

On the left of figure \ref{fig:bif}, two different bifurcation scenarios in
the domain where branch 2 emerges from 1 are drawn schematically. We see that
the part of branch 2 with an inward dimpled south pole emerges continuously
from the spherical branch 
1 and first runs to the right, i.e.\ to higher capsule
volumes.
After a turning point, the branch
runs to the left, i.e.\ to lower capsule volumes, while the south pole buckles
more and more inwards. 
The upper part of branch 2 up to the turning point (shaded dark red)
lies at higher energies than the spherical 
branch and consists of {\em unstable} stationary shapes.
This can be seen from the $p(V)$ diagram where 
it  corresponds to the lower part of branch 2 beyond the 
vertical turning point
and which are 
 unstable with respect to a volume-preserving  deformation mode
according to the generalization of criterion (iii)  \cite{Maddox87}.

Branch 2 intersects the spherical branch 1 at the critical volume 
$V=V_c$. For $V<V_c$ the lower part of branch 2 is the
energetically preferable solution branch. A shape transition between 
 branches 1 and 2 at $V_c$ is {\em discontinuous}. 
If the capsule wants to switch from the metastable branch 1 to 2
for $V<V_c$ (vertical arrow)  an energy barrier must be
overcome.
The upper part of  branch 2  represents 
 the unstable transition states  between a spherical shape 
 and the stable lower part of branch 2. Therefore, the energy barrier 
can be estimated by the energy difference between the upper 
and lower parts of branch 2.

The behavior when branch 2' emerges from the spherical branch 1
at the volume $V=V_{cb}$  
is quite different. It emerges with
egg-like configurations {\em continuously}
 from the trivial branch and runs directly
to the left. If the capsule passes this point in the bifurcation diagram
during a progressive reduction of its volume along the metastable branch 1,
 it is allowed to switch from
branch 1 to 2' continuously. Thus, there is no energy barrier to be overcome
in this scenario, and the trivial branch is supposed to be unstable.

Details of a realistic buckling process for given pressure can be constructed
from the $p(V)$ and $G(p)$ diagrams. The middle
and right parts of figure
\ref{fig:bif} schematically 
show the key features concerning the simply buckled states 2 and $2_c$. 
The decreasing  part of branch 2 with $\diff  p / \diff  V < 0$
is mechanically unstable with respect to volume reduction 
(shaded light red).
According to the generalization of criterion (iii) 
based on Ref.\ \cite{Maddox87} also the lower part of branch 2
(shaded dark red),
which has a positive slope for 
 a small volume range $V> V_{cb}$ in the $p(V)$ diagram, 
is unstable with respect to a volume-preserving  deformation mode.
It corresponds to the unstable upper branch 2 in the $F(V)$ 
diagram.
In the $G(p)$ diagram both corresponding unstable parts of branch 2 
join to give  the energetically unfavorable upper branch.  

Therefore, the spherical shapes of branch 1 become mechanically unstable
at the classical buckling pressure $p_{cb}$, where the unstable 
branch  2 and  branch 1 
 merge in the $p(V)$ diagram. Then,
a small dimple   caused by
fluctuations or agitations can grow spontaneously and the 
capsule  finally ends up in a fully collapsed
stable configuration $2_c$ at the given pressure running along the 
dashed path in the $p(V)$ diagram. 
The same process is indicated  in a
schematic $G(p)$ diagram on the right of figure
\ref{fig:bif} as a dashed line.

Shape $2_c$ becomes energetically preferable already at a much smaller 
negative pressure $p_c$, i.e.,  $|p_c|<|p_{cb}|$, where 
branches 1 and $2_c$ intersect in the  $G(p)$ diagram.
 A shape transition between 
 branches 1 and $2_c$  at $p_c$ is {\em discontinuous}:
Changing onto branch $2_c$  at $p=p_c$ does 
\emph{not} correspond to a spontaneous  snap-through 
into a fully  buckled shape $2_c$ because shape 
$1$ remains mechanically (meta)stable  in 
that region, because  $\diff  p / \diff  V > 0$ 
and no other branches are merging/intersecting 
 in  the  $p(V)$ diagram at  $p_c$. 
 A \emph{finite} dimple has to form by
fluctuations or agitations to induce buckling, and this is associated
with an energy barrier.
The unstable transition state of this process can be the 
unstable  shape 2 at the same pressure. 
This process is shown as a solid path in the $p(V)$ diagram. 
We note that, as a consequence of the 
condition of equal enthalpies $G$ at $p=p_c$, this solid path 
can be obtained by a Maxwell construction:
 At $p=p_c$, the two regions shaded gray in the $p(V)$ diagram have
   the same area.

We conclude this section by providing estimates for 
 buckling pressures for some synthetic and biological capsules.
For synthetic  capsules made from typical soft materials 
we expect  a Young's modulus in the range $E \sim 100-1000\,{\rm MPa}$, 
thicknesses $H_0 \sim 10-50\,{\rm nm}$, and  micrometer sizes 
$R_0 \sim 500\,{\rm nm}$, see for example 
 Refs.\ \cite{Zoldesi2008,Fery2004} for different synthetic capsules. 
This results in typical classical buckling pressures
 $|p_{cb}| \sim 0.1-1\, {\rm GPa}$
(in accordance with measurements in \cite{Fery2004}). 
Such materials have a small dimensionless bending modulus
$\tilde E_B \sim 0.00005 - 0.001$ corresponding to thin shells,
see  figures \ref{fig:001_L1}, \ref{fig:001_L2}, and \ref{fig:001_L3}. 

Many biological materials, such as virus capsids have very similar 
material characteristics but can be smaller: In 
Ref.\ \cite{Ru2009} $E =1 \, {\rm GPa}$, 
$H_0 = 2\,{\rm nm}$, and $R_0  = 10 \,{\rm nm}$ has been used for virus
capsids, which gives  similar dimensionless bending modulus
$\tilde E_B \sim 0.004$ and a  similar buckling threshold
$|p_{cb}| = 0.5\,{\rm GPa}$.

Somewhat different are soft biological capsules such as red blood cells with a 
shell made from lipid bilayers, which governs the bending rigidity. 
Red blood cells have a bending rigidity 
$E_B \sim 10\,k_BT$, an area stretching modulus $K \sim E H_0  \sim 10\,{\rm
  \mu N/m}$ \cite{Park2011}, and  sizes $R_0 \sim 4\,{\mu m}$, 
which results in a smaller 
 dimensionless bending modulus
$\tilde E_B \sim 0.0005$ and a much smaller 
classical buckling threshold $|p_{cb}| \sim 0.2\, {\rm Pa}$ in accordance with 
the fact that red blood cell shapes are buckled at ambient conditions. 

From our above result $p_c \approx  0.23 \, p_{cb}$
 for thin shells with $\tilde E_B \sim 0.001$, we expect 
 critical values $|p_c|$, which are smaller by a factor of at least $5$ 
compared to the classical buckling pressure 
$|p_{cb}|$ for all of  these capsules.

\section{Analysis of Simply Buckled Shapes and Bending Modulus}

\begin{figure}
 \centering
 \includegraphics[width=70mm]{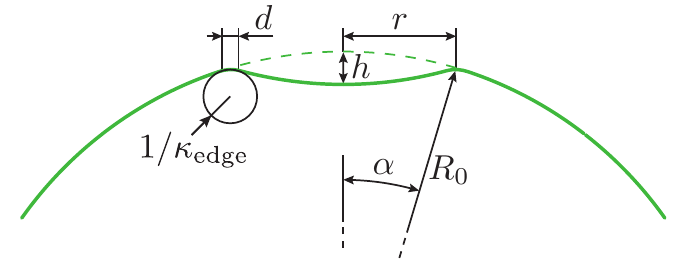}
 \caption{(Color online) Geometry of an axisymmetric dimple
   in a simply buckled shape 2.}
 \label{fig:dimple}
\end{figure}

Smaller bending resistances allow sharper bends in buckled
configurations. Hence, the minimal radius of curvature 
$1/\kappa_{\rm edge} \equiv  \min_s(1/\kappa_s)$ 
(in $s$-direction along the contour), 
which occurs close to the edge of the indentation of a 
simply buckled shape 2 (see figure \ref{fig:dimple}),
should depend sensitively on the reduced bending resistance
$\tilde E_B$ and represent an adequate observable to infer the 
reduced bending
modulus $\tilde E_B$.

\begin{figure}
 \centering
 \includegraphics[width=86mm]{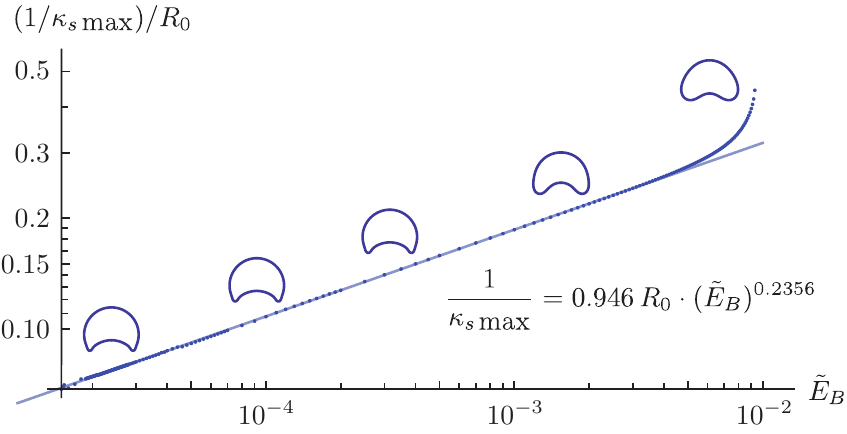}
 \caption{(Color online) Minimal radius of curvature 
   $1/\kappa_{\rm edge}$ plotted against the
   reduced bending resistance $\tilde E_B$. 
Dots: Computed shapes for $V=0.8 \, V_0$. Line: Power law fit for
   $\tilde E_B \leq 0.004$.}
 \label{fig:kappa_s}
\end{figure}

Figure \ref{fig:kappa_s} shows a double logarithmic plot of the 
minimal radius of
curvature as a function of the reduced bending modulus. It was obtained from a
series of simply buckled shapes with fixed volume $V = 0.8\, V_0$. 
For bending
moduli $\tilde E_B \leq 0.004$, a power law can be fitted to the data:
\begin{equation}
 \frac{1}{\kappa_{\rm edge}} \sim R_0\, {\tilde E_B}^{0.2356}.
\label{eq:fit}
\end{equation}

The power law (\ref{eq:fit}) can be confirmed by a scaling argument, 
where we balance bending and stretching energies. 
We consider a small dimple with radius $r\ll R_0$ and
depth $h$ as depicted in figure \ref{fig:dimple}. The elastic energy is mainly
located at the edge of this dimple, which spans the width $d\ll r$.

Since the dimple is assumed to be mirror-inverted to its original shape, its
radius and depth are determined by the angle $\alpha$ (see figure
\ref{fig:dimple}). To leading order, they are given by $r \sim R_0 \, \alpha$
and $h \sim R_0 \, \alpha^2$ respectively. By that, the volume change with
respect to the unbuckled configuration is approximately $\Delta V
\sim h\,r^2 \sim R_0^3 \, \alpha^4$.

The energies of bending and stretching are of the order
\begin{equation}
  F_b \sim E_B\, \frac{\zeta^2}{d^4}\,r\,d \quad \text{and} \quad 
   F_s \sim EH_0 \, \frac{\zeta^2}{R_0^2} \, r\,d
\end{equation} 
respectively \cite{LL7}, where $\zeta$ is the typical radial displacement of
the membrane near the edge of the dimple. Because the direction of the
meridian changes about $\alpha$ within the width $d$, we have $\zeta \sim
\alpha\,d \sim d \, (\Delta V/R_0^3)^{1/4}$. The total elastic energy for a  
given volume reduction $\Delta V$ therefore takes the form
\begin{equation}  
 F \sim EH_0 \frac{d^3\,r\,\Delta V^{1/2}}{R_0^{7/2}} + 
  E_B \frac{r\,\Delta V^{1/2}}{d\,R_0^{3/2}}.
\end{equation}
Minimizing the total elastic energy  with respect to $d$ we find 
the equilibrium  width of the dimple edge,
\begin{equation}
 d \sim \left( \frac{E_B}{EH_0} R_0^2 \right)^{1/4} \sim 
    R_0 \, \tilde E_B^{1/4},
\end{equation}
which can also be written in the form $d\sim\sqrt{H_0 \, R_0}$
 as in ref.\ \cite{LL7}.
Confining the directional change $\alpha$ of
the meridian to a width $d$ of the edge of the 
indentation (see figure \ref{fig:dimple}) results in an
edge  curvature
\begin{equation}
 \kappa_\text{edge} \sim \frac{\alpha}{d} \sim 
   \frac{\Delta V^{1/4}}{R_0^{7/4}} \, \tilde E_B^{-1/4}. 
\label{eq:scaling}
\end{equation}
 When the critical buckling pressure
$|p_c|$ is small (compared to the pressure unit defined by $EH_0/R_0$), 
the unbuckled region 
outside the dimple remains roughly 
spherical with a radius close to the original radius $R_0$, as 
can be seen from  eq.\ \ref{Rp1}. Therefore,  
\begin{equation}
  \Delta V = V_0-V = (1-v)V_0
\end{equation}
holds to a good approximation, where $v \equiv V/V_0$ is a reduced volume.
Using this in (\ref{eq:scaling}), we find  a scaling law
\begin{equation}
 \kappa_\text{edge} \sim 
   \frac{(1-v)^{1/4}}{R_0} \, \tilde E_B^{-1/4},
\label{eq:scaling2}
\end{equation}
which is  of the same form  as the above fit (\ref{eq:fit}) with
an  exponent $1/4$ matching the fit result $0.2356$ from
(\ref{eq:fit}) quite well.

It is evident from the assumption $d\ll r$, which corresponds to a sharp edge
of the indentation, that the scaling law holds only for sufficiently small
bending resistances, $\tilde E_B\leq 0.004$ in this case. The assumption $r\ll
R_0$ implies that the scaling law holds for sufficiently small volume
changes. Indeed, analyzing the scaling behavior for several capsule volumes 
using (\ref{eq:scaling2}), we find that the fitted exponent matches
the theoretical value $1/4$ very well for $V = 0.9 \, V_0$ or $0.8 \, V_0$,
but starts to deviate 
for $V = 0.67\,V_0$ or $0.5\, V_0$, see table \ref{table:fits}.
We can also determine that the numerical prefactor in (\ref{eq:scaling2})
is of the order of unity
and  only weakly volume dependent.
In contrast to these findings for simply buckled shapes 2 
without contact of opposite sides, we observe that $\kappa_\text{edge}$ 
is nearly {\em independent} of $\tilde E_B$ for buckled conformations
 of branch $2_c$ with opposite sides in contact.

\begin{table}[t]
\begin{center}
\begin{tabular}{|c|c|c|}
        \hline
  reduced Volume $v$ & prefactor $c$ & exponent $b$ \\ \hline
  $0.9$ & 0.73 & 0.246 \\
  $0.8$ & 0.63 & 0.236 \\
  $0.75$ & 0.59 & 0.229 \\
  $0.67$ & 0.54 & 0.221 \\
  $0.5$ & 0.50 & 0.217 \\
        \hline
\end{tabular}
\caption{\label{table:fits} Fit parameters for different capsule volumes. The
  fit model for the radius of curvature at the dimple edge is
  $1/\kappa_\text{edge} = c \, {(1-v)^{-1/4}} \, R_0 \, {\tilde E_B}^{b}$,
see eq.\ (\ref{eq:scaling2}).
 For small volume changes, the theoretical exponent $1/4$ agrees
  best.}
\end{center}
\end{table}

The results of the fits presented in table \ref{table:fits} could
be used to quantitatively  analyze experimental 
shapes of simply buckled  elastic
capsules without opposite sides in contact 
as shown, for example, 
in Refs.\ \cite{Fery2004,Zoldesi2008,Sacanna2011,Quilliet2008}, 
provided the radius of curvature at
the edge of the dimple can be measured accurately.
We note that the calculated
 shapes as shown in figure \ref{fig:kappa_s} are in 
qualitative agreement with some of the experimentally observed shapes
 \cite{Fery2004,Zoldesi2008,Sacanna2011,Quilliet2008}.
From a measurement of the curvature at the edge 
of the buckling dimple $\kappa_{\rm edge}$ the 
 dimensionless bending modulus  $\tilde E_B =
E_B / (R_0^2 \, E\,H_0)$ can be determined using (\ref{eq:scaling}) 
and the numerical prefactor from  table \ref{table:fits}.
In combination with an independent measurement of Young's modulus $E$,
for example, via  a measurement of the classical buckling pressure $p_{cb}$, 
this type of  shape analysis provides a method to obtain the bending modulus
of a capsule, which is hard to measure otherwise.

\section{Conclusions}

We applied nonlinear shell theory to the problem of axisymmetric deformations
of an initially spherical capsule. The elastic properties of the capsule
membrane were modeled with a quadratic strain-energy function. This approach
to Hooke's law allowed us to use the methods of force and moment equilibrium
and a least-energy principle simultaneously.

Bifurcation diagrams for reduced capsule volume and negative pressure were
presented. The least-energy principle gave information about the preferred
configurations and allowed to obtain a complete picture of the transition into
the fully buckled state. If the capsule volume is controlled, simply buckled
configurations turned out to be energetically preferable below a certain
critical volume $V_c$, but an energy barrier must be overcome in order to
leave the trivial solution branch. The transition at $V_c$ is thus
discontinuous.  If the internal negative pressure is controlled, spherical
shapes become mechanically unstable at the classical buckling pressure
$p_{cb}$.  However, buckled shapes with opposite sides in contact become
energetically favorable at a much lower negative pressure $p_c$, i.e.,
$|p_c|<|p_{cb}|$.  Also at controlled negative pressure, the transition at
$p_c$ is discontinuous and involves an energy barrier.
With the methods presented here, also 
configurations with opposite sides in contact could be computed and
incorporated in the bifurcation diagrams;  fully buckled 
configurations
with opposite sides in contact (branch $2_c$ in the bifurcation 
diagrams) actually have the lowest energies 
at small volumes or low negative pressures and determine the 
critical pressure  $p_c$.

For buckled shapes, 
the maximal  curvature $\kappa_\text{edge}$
at the edge of inward buckled dimples was found to
depend depend strongly 
 on the ratio of bending resistance to surface Young modulus,
with smaller ratios leading to sharper bends. A power law 
 $\kappa_\text{edge} \propto (EH_0/E_B)^{1/4}$ was found for
sufficiently small bending resistances and sufficiently small volume
changes. This relation may be used to analyze experimental 
shapes of buckled  elastic
capsules and  extract the bending  modulus of the capsule membrane from 
the capsule shape.

\begin{acknowledgments}
We thank Martin Brinkmann for discussions and bringing Ref.\ 
\cite{Maddox87} to our attention. 
  We acknowledge financial support by the 
 Mercator Research Center Ruhr (MERCUR).

\end{acknowledgments}

\appendix

\section{Calculus of Variations}
 \label{appendix_variation}

In this appendix we derive the first variation and the resulting 
Euler-Lagrange equations for  the 
  enthalpy functional 
\begin{equation}
 G = \int_0^{L_0} \left( 2 \pi\, r_0\,w_S  
 - p \, \pi \, r^2\,\lambda_s\,\sin\psi\right) \; \diff s_0.
\end{equation} 
The integrand has to be regarded as a function of
$s_0$. In functions like $r(s)$, a change of variables from $s$ to $s_0$ can
be performed by the function $s(s_0)$ introduced in (\ref{eq:def_lambda}).

Two of the functions $r(s_0)$, $z(s_0)$ and $\psi(s_0)$ determine the capsule
configuration completely. As the integrand of (\ref{eq:sphere_functional}) and
the geometrical relations contain mainly $r$ and $\psi$, we choose $r(s_0)$
and $\psi(s_0)$ as the two basic fields. 
Variations  $\delta
r(s_0)$ and $\delta \psi(s_0)$ have to fulfill  the boundary
conditions
\begin{equation}
  \delta r(0) = \delta r(L_0) = 0 \quad \text{and} \quad 
 \delta \psi(0) = \delta \psi(L_0) = 0.
\end{equation} 
The first variation  $\delta G$  of $G[r,\,\psi]$ is obtained as 
\begin{equation}
 \delta G = \int_0^{L_0} \left( 2 \pi\, r_0\,\delta w_S  
 - p \, \pi \, \delta(r^2\,\lambda_s\,\sin\psi)\right) \; 
 \diff s_0.
 \label{eq:variation_sphere}
\end{equation}
 The variation $\delta w_S$ of the surface
energy density introduces tensions and bending moments 
with the help of the constitutive equations
(\ref{stress-strain_2D}),
\begin{align}
 \delta w_S &= \frac{\del w_S}{\del e_s}\, \delta e_s 
 + \frac{\del w_S}{\del e_\phi}\, \delta e_\phi 
  + \frac{\del w_S}{\del K_s}\, \delta K_s 
   + \frac{\del w_S}{\del K_\phi}\, \delta K_\phi \nonumber \\
 &= \lambda_\phi\,\tau_s \, \delta e_s 
  + \lambda_s\,\tau_\phi\, \delta e_\phi 
   +  \lambda_\phi\,m_s\, \delta K_s 
  +  \lambda_s\,m_\phi\, \delta K_\phi. 
 \label{eq:sphere_vari_wS}
\end{align}
Now, the variations of the strains must be expressed in terms of $\delta r$,
$\delta \psi$ and its derivatives with the help of strain definitions and
geometrical relations,
\begin{equation}
 \begin{aligned}
 e_s &= \frac{r'(s_0)}{\cos \psi(s_0)} - 1
   & &\Rightarrow& \delta e_s 
  &= \frac{\delta r'}{\cos \psi} + \lambda_s \, \tan \psi \, \delta \psi \\
 e_\phi &=  \frac{r}{r_0} - 1 
  & &\Rightarrow& \delta e_\phi 
  &= \frac{1}{r_0}\, \delta r \\
 K_s &= \psi'(s_0) - \kappa_{s_0} 
  & &\Rightarrow& \delta K_s 
  &= \delta \psi' \\
 K_\phi &= \frac{\sin \psi}{r_0} - \kappa_{\phi_0} 
  & &\Rightarrow& \delta K_\phi 
  &= \frac{\cos \psi}{r_0} \, \delta \psi.
\end{aligned}
\label{eq:sphere_strain_vari}
\end{equation}
In a similar fashion, the variation of the second term can be calculated as
\begin{multline}
 \delta(r^2\, \lambda_s\,\sin \psi) = \delta(r'\,r^2\,\tan\psi) \\
 = 2\,r\,\lambda_s\,\sin\psi \, \delta r
 + r^2\,\tan\psi\,\delta r'
 + \frac{\lambda_s\,r^2}{\cos \psi} \, \delta \psi.
\end{multline}
Inserting everything into (\ref{eq:variation_sphere}), and sorting according
to $\delta r$, $\delta r'$, $\delta \psi$, $\delta \psi'$ yields a rather long
expression. Using integration by parts to transform $\delta r'$ into $\delta
r$ and $\delta \psi'$ into $\delta \psi$ (note that the boundary terms vanish)
results in the following result for the first variation of $G$, 
\begin{multline}
  \delta G = \int_0^{L_0} \!\!\!\! \diff s_0 
   \Big( \delta r \Big\{ 2\pi\,\lambda_s\,\tau_\phi 
    - 2\pi\,p\,r\,\lambda_s\,\sin\psi \\
    - 2\pi\,\frac{\diff}{\diff s_0}
    \!\!\Big( \frac{r\,\tau_s}{\cos\psi}\Big) 
   + \pi\,p\,\frac{\diff}{\diff s_0}(r^2\,\tan\psi) \Big\}  \\
   + \delta \psi \Big\{ 2\pi\,r\,\tau_s\,\lambda_s\,\tan\psi +
      2\pi\,\lambda_s\,m_\phi\,\cos\psi \\
      - \pi\,p\,\frac{\lambda_s\,r^2}{\cos\psi} -
      2\pi\,\frac{\diff(r\,m_s)}{\diff s_0} \Big\}\Big). 
\end{multline}
For a stationary shape, $\delta G=0$ for arbitrary  variations  $\delta \psi$
and $\delta r$, and the terms in curly braces have to vanish.
This gives 
the Euler-Lagrange equations describing stationary states.
 Rearranging the term next to $\delta \psi$ by a
change of variables $\diff s = \lambda_s \, \diff s_0$, we obtain
\begin{align}
0 &= \frac{\cos \psi}{r} \, m_\phi - 
  \frac{1}{r}\,\frac{\diff(r\,m_s)}{\diff s} - q \\
\text{with} \quad q &= -\tau_s \, \tan \psi +
    \frac{1}{2}\,p\,\frac{r}{\cos\psi}. 
\end{align}
which are  eqs.\ (\ref{eq:EL_1}) and (\ref{eq:sphere_q})
in the main text. 
With this definition of the transverse shear tension $q$, the term next to
$\delta r$ can be simplified to
\begin{equation}
 0 = \frac{1}{r}\,\frac{\diff(r\,\tau_s)}{\diff s} 
- \frac{\cos \psi}{r} \, \tau_\phi - \kappa_s \, q. 
\label{eq:sphere_equil_2}
\end{equation}
which gives eq.\ (\ref{eq:EL_2}) in the main text.

%

\end{document}